# Charge-4$e$ and charge-6$e$ flux quantization and higher charge superconductivity in kagome superconductor ring devices

Jun Ge[1]*, Pinyuan Wang[1]*, Ying Xing[1,2], Qiangwei Yin[3], Anqi Wang[4,5], Jie Shen[4,6], Hechang Lei[3], Ziqiang Wang[7] & Jian Wang[1,8,9†]

[1]*International Center for Quantum Materials, School of Physics, Peking University, Beijing 100871, China*

[2]*State Key Laboratory of Heavy Oil Processing, College of New Energy and Materials, China University of Petroleum, Beijing 102249, China*

[3]*Beijing Key Laboratory of Optoelectronic Functional Materials & Micro-Nano Devices, Department of Physics, Renmin University of China, Beijing 100872, China*

[4]*Beijing National Laboratory for Condensed Matter Physics, Institute of Physics, Chinese Academy of Sciences, Beijing 100190, China*

[5]*School of Physical Sciences, University of Chinese Academy of Sciences, Beijing 100049, China*

[6]*Songshan Lake Materials Laboratory, Dongguan 523808, China*

[7]*Department of Physics, Boston College, Chestnut Hill, MA 0246, USA*

[8]*Collaborative Innovation Center of Quantum Matter, Beijing 100871, China*

[9]*Hefei National Laboratory, Hefei 230088, China*

[†]To whom correspondence should be addressed. jianwangphysics@pku.edu.cn (J.W.).

* These authors contribute equally.

**The flux quantization is a key indication of electron pairing in superconductors. For example, the well-known $h/2e$ flux quantization is considered strong evidence for the existence of the charge-2$e$, two-electron Cooper pairs. Here we report evidence for multi-charge flux quantization in mesoscopic ring devices fabricated using the transition-metal kagome superconductor $CsV_3Sb_5$. We perform systematic magneto-transport measurements and observe unprecedented quantization of magnetic flux in units of $h/4e$ and $h/6e$ in magnetoresistance oscillations. Specifically, at low temperatures, magnetoresistance oscillations with period $h/2e$ are detected, as expected from the flux quantization for charge-2$e$ superconductivity. We find that the $h/2e$ oscillations are suppressed and replaced by resistance oscillations**

**with $h/4e$ periodicity when temperature is increased. Increasing the temperature further suppresses the $h/4e$ oscillations and robust resistance oscillations with $h/6e$ periodicity emerge as evidence for charge-6$e$ flux quantization. Our observations provide the first experimental evidence for the existence of multi-charge flux quanta and emergent quantum matter exhibiting higher-charge superconductivity in the strongly fluctuating region above the charge-2$e$ Cooper pair condensate, revealing new insights into the intertwined and vestigial electronic order in kagome superconductors.**

## I. INTRODUCTION

Superconductivity was discovered more than one century ago and described by the Bardeen–Cooper–Schrieffer (BCS) theory in terms of the condensation of charge-2$e$ Cooper pairs [1,2]. Flux quantization in units of the charge-2$e$ flux quantum $h/2e$, where $h$ is the Planck constant and $e$ is the elementary charge, is a key signature of the electron pairing in BCS superconductors. The observations of the $h/2e$ flux quantization served as key experimental evidence of the BCS theory of conventional superconductors (SCs) [3–5]. Superconductivity via the condensation of higher charges, such as bound states of electron quartets or sextets (i.e. charge-4$e$ or charge-6$e$), has been proposed theoretically [6-16]. Despite intensive theoretical studies, whether higher-charge superconductivity exists and the many of its basic properties remain mysterious and illusive due to the lack of experimental evidence. For charge-Q superconductivity with Q=2n$e$ to arise, there must exist experimental evidence for flux quantization in units of the charge-Q flux quantum $\Phi_0^Q = h/2ne$.

Here, we report the observation of flux quantization in units of $\Phi_0^{4e}$ ($h/4e$) and $\Phi_0^{6e}$ ($h/6e$) in ring devices of the newly emerged kagome superconductor $CsV_3Sb_5$. We fabricate superconducting $CsV_3Sb_5$ ring devices with various sizes and perform systematic magneto-transport measurements. At low temperatures, magnetoresistance oscillations with period of the charge-2$e$ flux quantum $h/2e$ are detected, as expected from the flux quantization for charge-2$e$ superconductivity. The $h/2e$ oscillations are suppressed and replaced by resistance oscillations with $h/4e$ periodicity as the temperature is increased. When the temperature is further increased toward the onset of the superconductivity, the $h/4e$ oscillations vanish and novel resistance oscillations with a period equaling to 1/3 of the period of $h/2e$ oscillations emerge. Careful measurements and analysis on many ring devices indicate consistently that these resistance oscillations with the period equaling to 1/2 and 1/3 of the $h/2e$ oscillations reveal the existence of flux quantization in units of charge-4$e$ and charge-6$e$ (flux quanta $\Phi_0^{4e}$ and $\Phi_0^{6e}$), respectively. Our observations provide the first and direct experimental evidence of flux quantization in units of the multi-charge flux quanta in kagome superconductor ring devices and point to the possible existence of unprecedented higher-charge superconducting quantum states in the strongly fluctuating region of kagome superconductors.

## II. RESULTS

Bulk $CsV_3Sb_5$ crystallizes into the P6/mmm space group [Fig. 1(a)], with a two-dimensional (2D) kagome network of vanadium (V) cations coordinated by octahedra of antimony (Sb) [17,18]. Adjacent kagome sheets are separated by layers of cesium (Cs) ions [17,18]. Bulk $CsV_3Sb_5$ undergoes charge density wave transition at around ~ 94 K before developing superconductivity [18] at about 2.5 K. Due to the combined effects of geometric frustration, quantum interference and electron correlation, and electron-lattice interaction on the kagome network, a diverse set of correlated and topological electronic states has been discovered, including superconductivity and pair density wave order [13,19-32]. In this work, we fabricate the $CsV_3Sb_5$ ring structures by etching the kagome superconductor thin flakes exfoliated from bulk samples as shown in the schematic Fig. 1(b). Figure 2(a) displays the temperature dependence of the resistance of $CsV_3Sb_5$ ring device s1 with a thickness of ~ 14.6 nm in zero magnetic field, with the false-colored scanning electron microscope (SEM) image shown in the inset and the atomic force microscopy results in Fig. 5(a). The superconducting transition with the onset temperature $T_c^{onset}$ ~ 3.9 K and the zero resistance temperature $T_c^{zero}$ ~1.1 K marked in the figure delineates an extended superconducting fluctuation region where new quantum states of paired matter may arise.

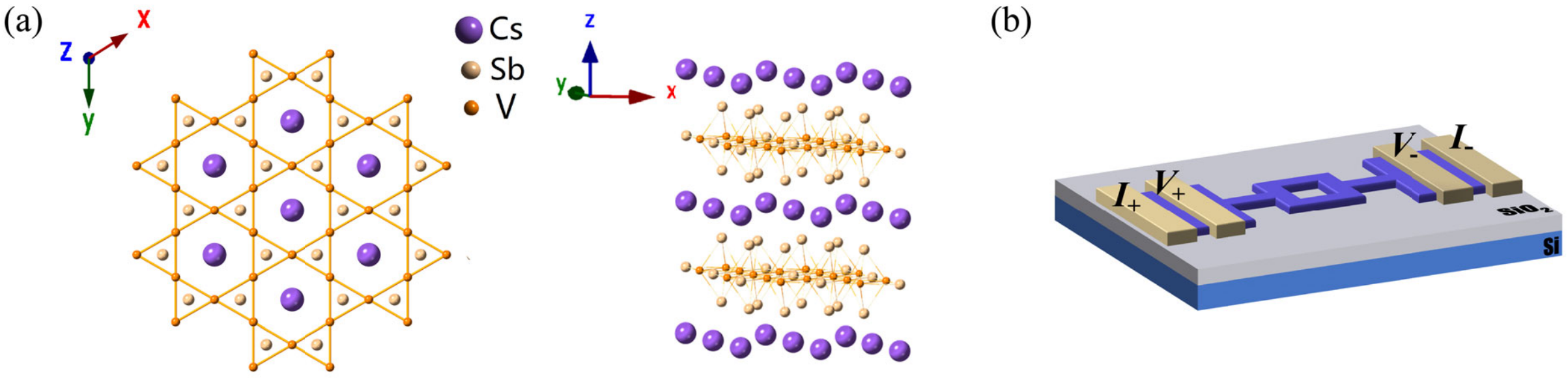

FIG. 1. Schematic crystal structure and drawing of the ring-structure device of $CsV_3Sb_5$. (a) Schematic crystal structure of $CsV_3Sb_5$ with purple, orange and yellow spheres denoting Cs, V and Sb atoms, respectively. (b) Schematic drawing of the ring-structure device in the standard four-terminal configuration. The dimensions are not drawn to scale.

We then carry out systematic magneto-transport measurements on the $CsV_3Sb_5$ ring devices. When the applied perpendicular magnetic field drives the sample into a sufficiently resistive state, noticeable oscillations in a magnetic field range from 2000 Oe to 7000 Oe are observed on the measured resistance ($R$)-magnetic field ($H$) curves shown in Fig. 7(a) at temperatures from 0.1 K to 1.0 K. By subtracting the smoothly rising (in $H$) resistance backgrounds in the $R$-$H$ curves, the oscillations become more pronounced as shown in Fig. 2(b) at various temperatures, where dashed and solid lines guide the dips and peaks, respectively. We index the dips and peaks of the oscillations by an integer or half-integer n and the corresponding magnetic field by $H_n$. The index n is obtained by $H_n/\Delta H$, here, $\Delta H$ is the period of the oscillation. The positions of the dashed or solid line associated with an index n are chosen to make sure that most of the corresponding dips and peaks on different curves can be properly located. Plotting $H_n$ versus n reveals a clear linear dependence [Fig. 2(h)], indicating that these oscillations are periodic in $H$, as expected from the quantization of the magnetic field flux through the ring $\Phi = n\Phi_0$, where $n$ is an integer, in unit of the charge-2$e$ superconducting flux quantum $\Phi_0$. Using the measured period of the oscillations in Fig. 2(b), i.e. $\Delta H_{2e} \sim$ 750 Oe, and setting $\Phi_0=h/2e=S\cdot\Delta H_{2e}$, we obtain the effective ring area $S \approx$ 0.0276 μm$^2$, which is marked by the red rectangle in the inset of Fig. 2(a). Note that if the effective area were larger, the condition for flux quantization would imply Cooper pairs of fractional charged quasiparticles, which is unlikely for this system. We thus consider the oscillations with period of ~ 750 Oe as the charge-2$e$ ($h/2e$) oscillations with the effective area close to inner hole area. We will return to the physical implications in the discussion section.

Surprisingly, upon increasing the temperature to 1.65 K and above into the broad superconducting transition region, new oscillations in a magnetic field range from 2000 Oe to 5000 Oe are observed [Fig. 2(c)] with a different period in magnetic field $\Delta H \sim$ 350 Oe as confirmed by the linear relation between n and $H_n$ [Fig. 2(h)]. These emergent oscillations correspond to a magnetic flux through the ring $\Phi=S\cdot\Delta H=0.97\times10^{-15}$ Wb, using the same effective area $S \approx$ 0.0276 μm$^2$ identified above and marked in the inset of Fig. 2(a). Intriguingly, this value of the flux is very close to one-half of the charge-2$e$ flux quantum, i.e. the charge-4$e$ flux quantum $\Phi_0^{4e}=h/4e=1.03\times10^{-15}$ Wb.

More strikingly, we find that the oscillations with period of ~350 Oe vanish when the temperature is further increased. Above 2.3 K, brand new oscillations in magnetic field emerge with the period $\Delta H \sim$ 250 Oe [Figs. 2(d)-2(g)]. This value is remarkably 1/3 of that in the charge-2$e$ oscillations (~ 750 Oe) observed at low temperatures and corresponds to resistance oscillations due to flux quantization in units of an emergent charge-6$e$ flux quantum $\Phi_0^{6e}$=$h$/6$e$ =0.69×10$^{-15}$ Wb, under the same effective area ($S$ ≈ 0.0276 μm$^2$). Indeed, the charge-6$e$ resistance oscillations in Figs. 2(d)-2(g) are strong and robust over a wide field range [Figs. 8, 9], exhibiting magnetic flux quantization $\Phi = n\Phi_0^{6e}$ down to zero magnetic field with a resistance minimum at $n$=0 [Figs. 2(d), 2(e)]. In Fig. 2(i) and Fig. 11(a), the temperature evolution of the fast Fourier transform (FFT) of the resistance oscillations is presented in a waterfall plot. The positions of the FFT peaks show that the periodicity of the flux quantization in unit of the magnetic flux quantum changes from $\Phi_0$=$h$/2$e$ at low temperatures to $\Phi_0^{4e}$=$h$/4$e$ at higher temperatures, and then to $\Phi_0^{6e}$=$h$/6$e$ at still higher temperatures on approaching the onset of the superconducting transition. Similar results have also been observed consistently in other $CsV_3Sb_5$ ring devices of similar dimensions. The results obtained on another device are summarized in Fig. 12.

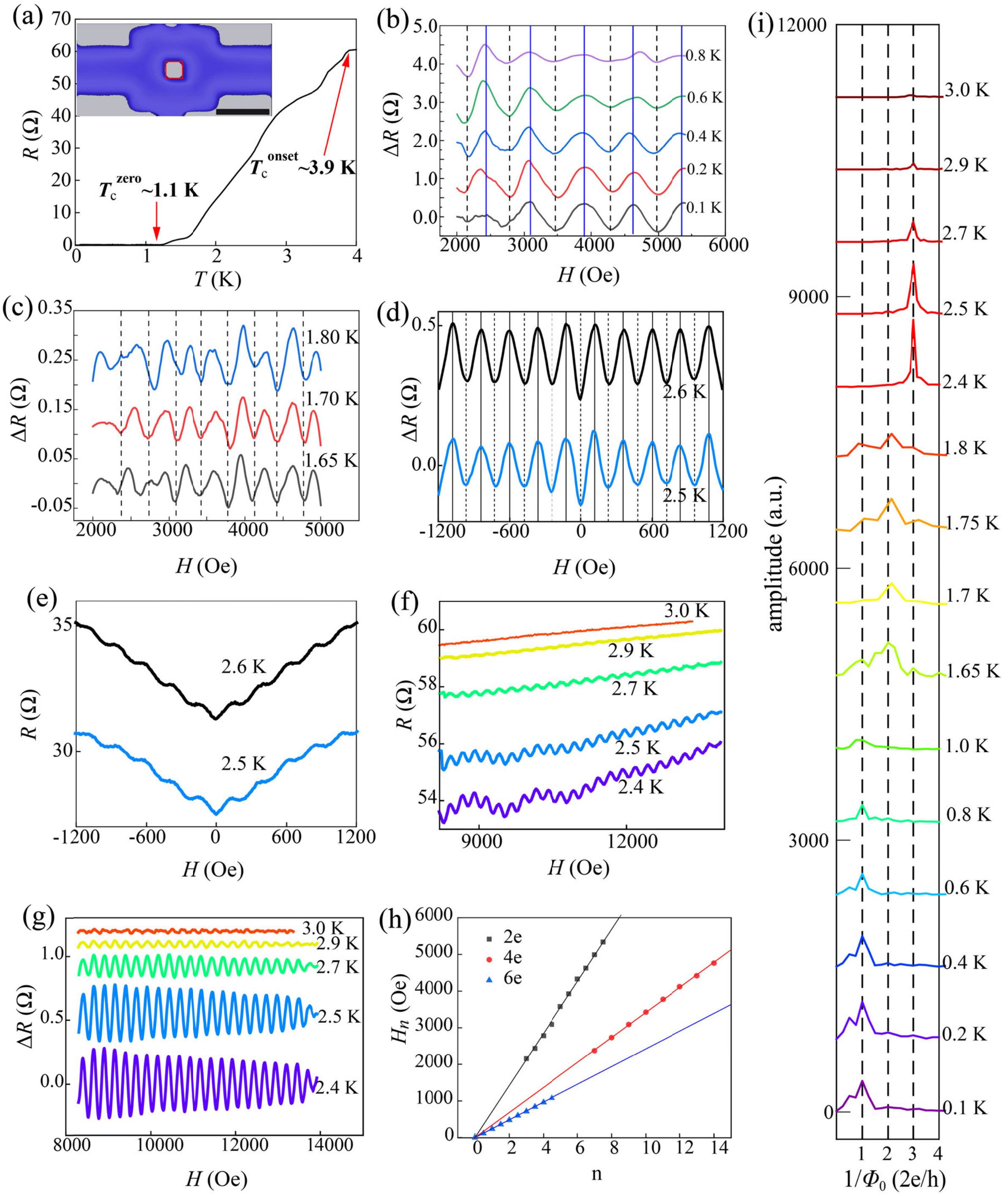

FIG. 2. Evolution of quantum oscillations in superconducting $CsV_3Sb_5$ ring device s1. (a) Resistance as a function of temperature from 4 K to 0.07 K. Superconductivity with onset temperature ~ 3.9 K and zero-resistance temperature ~1.1 K is observed. Inset shows the false-colored image of the $CsV_3Sb_5$ ring device s1. The sample protected by PMMA layer is represented by blue and the substrate is represented by gray. The scale bar in the false-colored image represents

500 nm. The inner area, middle area and outer area of the ring estimated from the false-colored image is ~0.027 μm$^2$, ~0.21 μm$^2$ and ~0.54 μm$^2$, respectively. (b) $h/2e$ oscillations with period of ~750 Oe in which the rising backgrounds have been subtracted as a function of the perpendicular magnetic field. The black dashed lines and the blue solid lines label the dips and peaks of the oscillations, respectively. For clarity, data curves are shifted. The effective area of the $h/2e$ oscillations (~0.0276 μm$^2$) is marked by the red rectangle in the inset of (a). (c) Oscillations with period of ~ 350 Oe at higher temperatures after subtracting smooth backgrounds. The dashed lines label the oscillation dips. For clarity, data curves are shifted. The period of the oscillations (~350 Oe) is nearly 1/2 of that of the $h/2e$ oscillations (~750 Oe). If we consider the effective area of the oscillations with period of ~350 Oe is the same as that of the $h/2e$ oscillations with period of ~750 Oe, then the period of ~350 Oe will correspond to a periodicity of $h/4e$. (d) Oscillations with period of ~250 Oe at still higher temperatures after subtracting smooth backgrounds. For clarity, data curves are shifted. The black dashed and solid lines label the dips and peaks of the oscillations, respectively. The period of the oscillations (~250 Oe) is 1/3 of that of the $h/2e$ oscillations (~750 Oe). If we consider the effective area of the oscillations with period of ~250 Oe is the same as that of the $h/2e$ oscillations with period of ~750 Oe, then the period of ~250 Oe will correspond to a periodicity of $h/6e$. (e) The raw data showing the oscillations with period of ~250 Oe in the low field region. (f), (g) Oscillations with period of ~250 Oe in the high field region. (h) n-$H_n$ index plots of the $h/2e$, $h/4e$ and $h/6e$ oscillations. Here, n is an integer or half integer. Integer n represents the oscillation dip and half integer n represents the oscillation peak. The linear relation between n and $H_n$ can be clearly observed, indicating that the $h/2e$, $h/4e$ and $h/6e$ oscillations are periodic. (i) FFT results as a function of inverse magnetic flux in unit of $2e/h$ at various temperatures. The FFT curves are scaled in amplitude for clarity. Note that the oscillation range for FFT is not exactly the same as the range shown in (b)-(d). With increasing temperature, the changes in periodicity from $h/2e$ to $h/4e$ and then to $h/6e$ can be clearly observed. Note that the weak peaks appearing around 0.5×($2e/h$) on FFT curves at $T < 0.6$ K correspond to the secondary peaks, which may originate from the slight amplitude fluctuation of the $h/2e$ oscillation signals.

It is important to note that these set of thick-rimmed ring devices leave room for the possibility that the oscillations with period of ~350 Oe and ~250 Oe were also $h/2e$ oscillations, but with larger effective areas. In the following, we will show that this possibility can be excluded and the effective area remains a constant for $h/2e$, $h/4e$ and $h/6e$ oscillations in a given $CsV_3Sb_5$ ring device. First, as shown in Fig. 2(i) and Fig. 11(a), the FFT peak positions for $h/2e$, $h/4e$, and $h/6e$ oscillations respectively do not change with the temperature, indicating that the corresponding effective area maintains a constant with increasing temperatures.

Second, there are no intermediate periodicities between the $h/2e$ ($h/4e$) and $h/4e$ ($h/6e$) oscillations, which further demonstrates the constant effective area for the observed oscillations.

To further scrutinize the evidence for the extraordinary charge-$6e$ flux quantization and directly exclude the possibility of assigning the $h/6e$ oscillations as those of $h/2e$ under a different effective area, we fabricated micron-sized $CsV_3Sb_5$ ring devices with much larger hole area and ratio of hole size to wall width. The experimental data on device s2 with a hole area ~ 0.96 μm$^2$ are presented in Fig. 3. At low temperatures, resistance oscillations with period ~ 13.6 Oe in magnetic field are observed. Using the measured period of the oscillations in Fig. 3(b), i.e. $\Delta H_{2e}$ ~ 13.6 Oe, and setting $\Phi_0=h/2e=S\cdot\Delta H_{2e}$, we obtain the effective area $S \approx 1.52$ μm$^2$, which is marked by the red rectangle in the inset of Fig. 3(a). If a larger effective area, e.g. with boundaries in the middle of the rim ($S_{middle}$ ~2.01 μm$^2$) were considered as the effective area, then the flux $\Phi= S_{middle}\cdot\Delta H=2.734\times10^{-15}$ Wb$\approx h/0.77e$, would lead to fractional charges, which is unlikely in this system. Therefore, the oscillations at low temperatures in the large ring device s2 are $h/2e$ oscillations.

When the temperature is increased to 2.5 K and above, novel emergent oscillations are again observed [Fig. 3(e)], as in the smaller, thick-rimmed devices. The periodicity of these new oscillations is confirmed by the linear relation between n and $H_n$ [Fig. 3(f)], which reveals a new period in magnetic field $\Delta H$ ~ 4.6 Oe, equaling to 1/3 of the period of the $h/2e$ oscillations (~13.6 Oe) observed at low temperatures. These small-period oscillations (~4.6 Oe) cannot be mistaken for $h/2e$ oscillations with a larger effective area in the micron-sized device s2, because if they were, the corresponding effective area (green rectangle in the inset of Fig. 3(a)) would be much larger than even the outer area of the ring device, which is physically impossible. Therefore, the oscillations with period of ~ 4.6 Oe are unambiguous evidence for flux quantization in units of novel higher-charge magnetic flux quantum. Considering the extremities of device s2, i.e. an outer area $S_{outer}$~2.97 μm$^2$ and an inner area $S_{inner}$~0.96 μm$^2$, the oscillations period ~4.6 Oe imply that the flux quantum must satisfy: $\Phi_{inner} \leqslant \Phi \leqslant \Phi_{outer}$, where $\Phi_{inner}= S_{inner}\cdot\Delta H=0.44\times10^{-15}$ Wb$\approx h/9e$ and $\Phi_{outer}= S_{outer}\cdot\Delta H=1.366\times10^{-15}$ Wb$\approx h/3e$. This means that the oscillations with period of ~4.6 Oe do not correspond to the $2e$ but the multi-charge of the flux quantum in the range of $3e$ to $9e$.

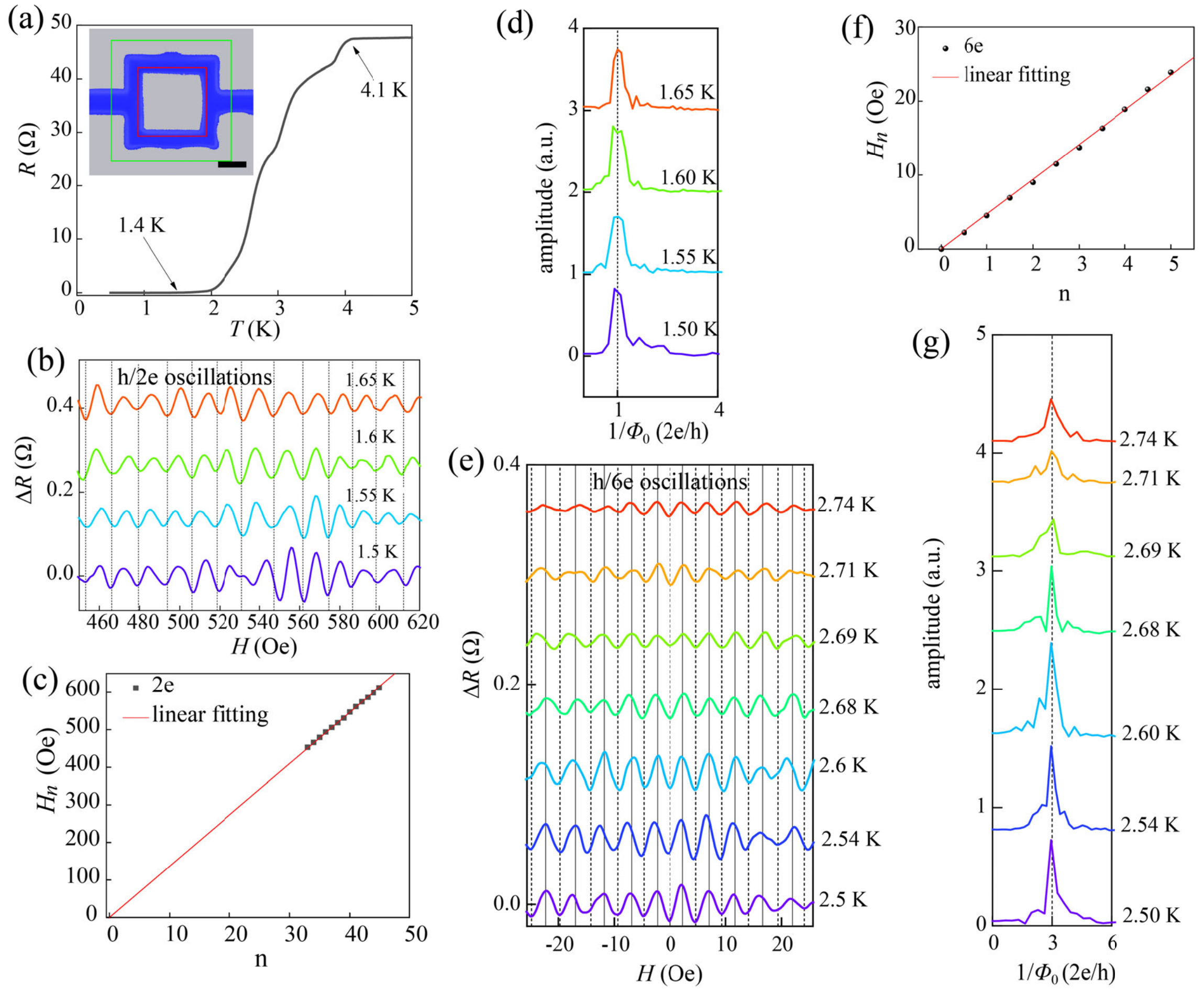

FIG. 3. $h/2e$ and $h/6e$ oscillations in the superconducting $CsV_3Sb_5$ ring device s2. (a) Resistance as a function of temperature from 5 K to 0.5 K. Superconductivity with onset temperature ~ 4.1 K and zero-resistance temperature ~1.4 K is observed. Inset shows the false-colored image of the $CsV_3Sb_5$ ring device s2. The sample protected by PMMA layer is represented by blue and the substrate is represented by gray. The scale bar in the false-colored image represents 500 nm. The inner area, middle area and outer area of the ring estimated from the false-colored image is ~0.96 μm$^2$, ~2.01 μm$^2$ and ~2.97 μm$^2$, respectively. (b) $h/2e$ oscillations with period of ~13.6 Oe after subtracting smooth backgrounds. The dashed lines label the oscillation dips. For clarity, data curves are shifted. The effective area of the $h/2e$ oscillations (~1.52 μm$^2$) is marked by the red rectangle in the inset of (a). (c) n-$H_n$ index plot of the $h/2e$ oscillations. Here, n is an integer, representing the oscillation dip. A linear relation between n and $H_n$ can be clearly observed, indicating that the magnetoresistance oscillations are periodic. (d) FFT results of the $h/2e$ oscillations. Note that the oscillation range for FFT is not exactly the same as the range shown in (b). (e) Oscillations with period of ~ 4.6 Oe

after subtracting smooth backgrounds. The dashed and solid lines label the oscillation dips and peaks, respectively. For clarity, data curves are shifted. The period of the oscillations (4.6 Oe) is nearly 1/3 of that of the $h/2e$ oscillations (13.6 Oe). If the oscillations with period of ~4.6 Oe were $h/2e$ oscillations, the effective area (~4.5 μm$^2$, shown by the green rectangle in the inset of (a)) would be obviously much larger than the outer area of the ring device (~2.97 μm$^2$), which is physically impossible. If we consider the effective area of the oscillations with period of ~4.6 Oe is the same as that of the $h/2e$ oscillations with period of ~13.6 Oe, then the period of ~4.6 Oe will correspond to a periodicity of $h/6e$. (f) n-$H_n$ index plot of the $h/6e$ oscillations. Here, n is an integer or half integer, representing the oscillation dip and peak, respectively. The linear relation between n and $H_n$ confirms that the magnetoresistance oscillations are periodic. (g) FFT results of the $h/6e$ oscillations. Note that the oscillation range for FFT is not exactly the same as the range shown in (e).

The evidence for a temperature-independent effective area discussed above encourages the use of the same effective area ($S\approx$1.52 μm$^2$) determined for the low-temperature charge-2$e$ oscillations with period ~13.6 Oe and the oscillations with period ~4.6 Oe at higher temperatures, which corresponds to a magnetic flux through the ring $\Phi=S\cdot\Delta H=0.699\times10^{-15}$ Wb. This value is remarkably close to the periodicity in unit of the charge-6$e$ superconducting flux quantum $\Phi_0^{6e}=h/6e$ $=0.69\times10^{-15}$ Wb. Similar results have also been obtained in two additional large ring devices s4 [Fig. 13] and s5 [Fig. 14]. These results on devices with much larger hole areas provide arresting evidence for the discovery of robust $h/6e$ oscillations in $CsV_3Sb_5$ ring devices. In Fig. 4, we present the temperature evolution of the magnetoresistance oscillations as color intensity plots in unit of the magnetic flux $\Phi/\Phi_0$ over the broad superconducting transition region in the $CsV_3Sb_5$ ring devices. With the increase of the temperature, the periodicity of the oscillations changes from $h/2e$ to $h/4e$ and then to $h/6e$ in small thick-rimmed ring devices represented by s1, and from $h/2e$ to $h/6e$ in micron sized ring devices represented by s2, recapitulating our finding of multi-charge flux quantization in unit of charge-4$e$ and charge-6$e$ superconducting flux quanta.

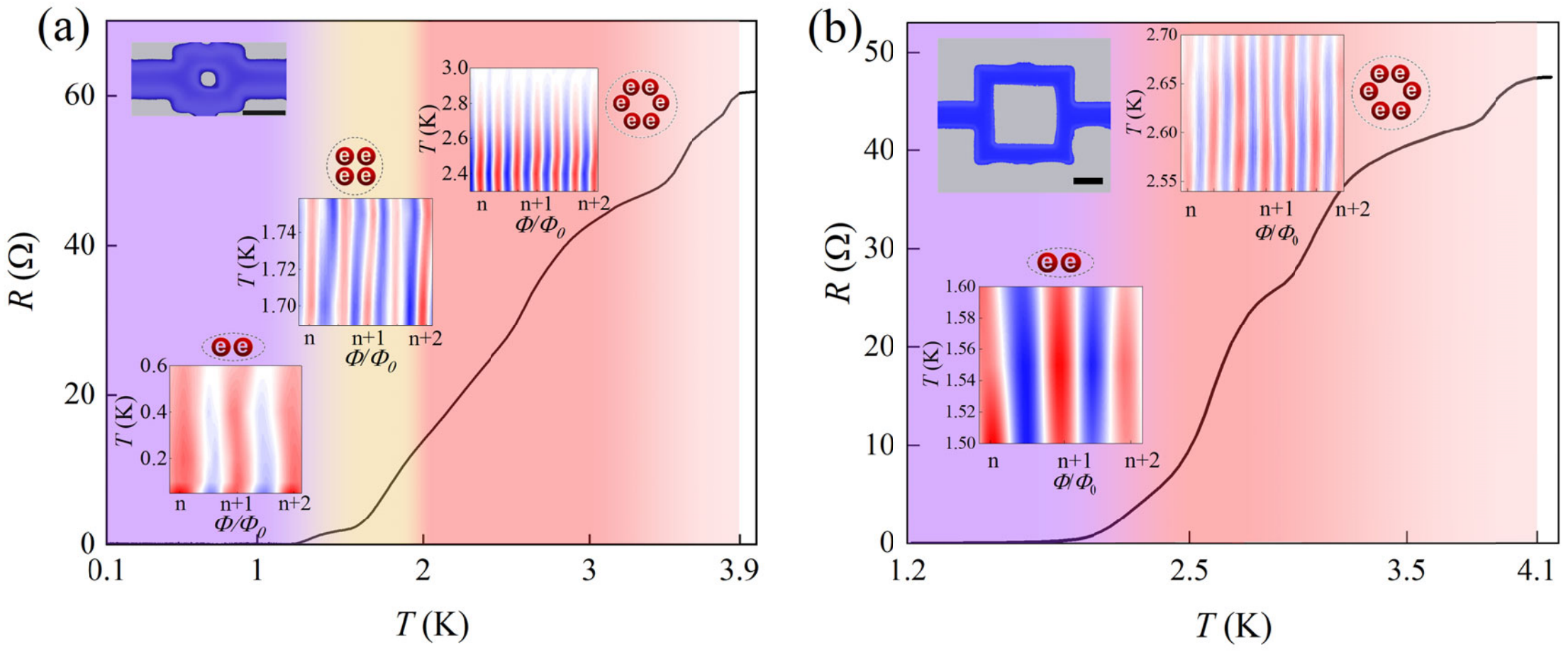

FIG. 4. *R*-*T* phase diagram of the *h*/2*e*, *h*/4*e* and *h*/6*e* flux quantization in $CsV_3Sb_5$ ring device s1 (inner area ~0.027μm$^2$) (a), and *h*/2*e* and *h*/6*e* flux quantization in $CsV_3Sb_5$ ring device s2 (inner area ~0.96 μm$^2$) (b). The upper left insets in (a) and (b) show the false-colored SEM images of $CsV_3Sb_5$ ring device s1 and s2, respectively, where the scale bars represent 500 nm. The dark solid line is the *R*-*T* curve in zero magnetic field. Different monochromatic color shaded regimes in different temperature ranges on the *R*-*T* plane, i.e. the purple region below 1.0 K in (a) and below 1.8 K in (b), the yellow region at 1.65 K<*T*<1.8 K in (a), the red region at 2.0 K<*T*<3.0 K in (a) and 2.5 K<*T*<2.74 K in (b), indicate the charge-2*e*, 4*e* and 6*e* phases identified by the corresponding 2*e*, 4*e*, and 6*e* flux quantization in magnetoresistance oscillations, respectively. The latter are presented as intensity maps (insets) of Δ*R* on the temperature *T* and magnetic flux $\Phi/\Phi_0$ plane, displaying the periodicity of the oscillations in the corresponding temperature range. The white region above 3.9 K in s1 and 4.1 K in s2 represents the normal state. The intermediate gradient ramp regions between the monochromatic color shaded regimes schematically illustrate the transition of different states. No periodic oscillations are observed in the intermediate gradient ramp region. With increasing temperature, the periodicity changes from *h*/2*e* to *h*/4*e* and then to *h*/6*e* in smaller ring device s1. In the large ring structure s2, the periodicity changes from *h*/2*e* to *h*/6*e*.

## III. DISCUSSION

We stress that these extraordinary flux quanta originate from the extraordinary superconducting fluctuation region in $CsV_3Sb_5$ ring devices that is very broad with internal structures as reflected by the *R*-*T* curve. The physics in this fluctuation region is different and beyond what happens in the transition region of typical BCS superconductors. For ordinary BCS superconductors, this transition region is narrow (the SC transition is sharp) and the melting of coherent charge-2*e* Cooper pairs is all that happens wherein, resulting in the canonical behavior of *h*/2*e* oscillations [33]. In contrast, the transition region in our kagome $CsV_3Sb_5$ ring devices is rather broad, as can be seen in the *R*-*T* curves. It separates the charge-2*e* SC ground state below the zero-resistance temperature and the normal state above the SC onset temperature. The multi-charge flux quanta emerge in the broad transition region with strong SC fluctuations, as the temperature is increased in the melting of the charge-2*e* condensate ground state. To further substantiate this fundamental difference and develop new insights, we studied ring devices made of conventional superconductor Nb [Figs. 15, 16] using identical fabrication methods. Two Nb ring devices are fabricated using the same etching technique, under the same etching parameters, and protected by the PMMA layer of the same thickness. The measured *R*-*T* curve reveals a sharp superconducting transition in the Nb ring device n1 [onset temperature ~7.6 K and zero resistance temperature ~7.0 K, Fig. 15(a)] and n2 [onset temperature ~7.6 K and zero resistance temperature ~7.2 K, Fig. 16(a)]. Indeed magnetoresistance measurements show only *h*/2*e* oscillations in Nb ring devices in a narrow temperature regime [Fig. 15(c), Fig. 16(d)] above the zero-resistance transition. The corresponding effective area has a radius lying in the middle of the wall [inset of Fig. 15(a) for device n1 and Fig. 16(a) for device n2], which is in accord with the theoretical predictions for the Little-Parks oscillations in conventional superconductor ring devices [34]. These experiments reassure that our device fabrication, measurements, and analysis have been tested to reproduce the expected physics in ordinary BCS superconductors, and further demonstrate that the remarkable observation of the multi-charge flux quantization is due to the never before encountered higher-charge superconducting quantum states in extended fluctuation region of $CsV_3Sb_5$ ring devices.

Another intriguing property, distinct from ordinary superconductors having sharp superconducting transitions, is that the observed resistance oscillations with different flux quantization in the strongly fluctuating superconductivity region all have the same effective area close to the inner hole area in the $CsV_3Sb_5$ ring devices [inset of Figs. 2(a), 3(a)]. In a recent insightful paper, Han and Lee [35] provided a possible theoretical explanation for our experimental observation. They studied the effective area for transporting charge-4*e* and charge-6*e* bound states across our thick-rimmed geometry devices ($CsV_3Sb_5$ ring device s1, s3) using the space-time formulation of time-dependent Ginzburg-Landau theory. Because the superconductivity is strongly fluctuating in the broad transition regime in the $CsV_3Sb_5$ ring devices, the optimal path is found to stick to the edge of the inner hole in order to reduce the fluctuations by proximity to the open area of the hole, leading to the effective area close to the inner area for *h*/4*e* and *h*/6*e* flux quantization. The result also applies to the strongly fluctuating charge-2$e$ state [35]. The combined experimental and theoretical findings land convincing support for the observation of resistance oscillations with $h/4e$ and $h/6e$ flux quantization.

We now turn to the possible origin of the observed charge-4$e$ and charge-6$e$ flux quantization. There are theoretical proposals for fractional flux quantum in spin-triplet $p + ip$ superconductors [36,37], and in vortices trapped at domain walls and twin boundaries in time-reversal symmetry breaking superconductors [38]. However, these do not describe our observation because the fractional flux quantization under these settings only leads to nontrivial phase shift in the quantum oscillations, while the periodicity of the oscillations would remain at $\Phi_0$ as the pairing is still charge-2$e$ in nature. More theoretical discussions are available in Appendix B. While we cannot rule out the possibility where a fractional flux quantum results from a multicomponent superconductor where the phases of different charge-2$e$ condensates wind differently in the magnetic field [39], the observed sequential changes in the flux quantization under thermal melting are unnatural to be accounted for in this scenario.

The distinct charge-4$e$ and charge-6$e$ flux quanta observed in the flux quantization with increasing temperatures naturally suggest a sequential destruction of the phase coherent charge-2$e$ superconductivity and the emergence of phase coherent bound states of 4-electrons (or two Cooper pairs) and 6-electrons (or three Cooper pairs) in the strongly fluctuating transition region of the ring structures. This scenario is consistent with the theoretical proposal of the putative charge-4$e$ and charge-6$e$ superconductivity as vestigial ordered states following the melting of a novel charge-2$e$ superconducting state that breaks crystalline symmetry [13] as evidenced by the PDW order observed in the kagome superconductor $CsV_3Sb_5$ [23]. In addition to the broken global U(1) symmetry, such a PDW superconductor has a multicomponent charge-2$e$ order parameter that simultaneously breaks translation and time-reversal symmetry, where the relative chiral phases of the different components creates an emergent hexagonal vortex-antivortex lattice carrying the momentum of the PDW. The destruction of the PDW order by thermal fluctuations is thus described by the well-known melting theory of a vortex-antivortex lattice [40-42]. With increasing temperatures, the topological defects of positional order, i.e. dislocations, unbind and proliferate, which destroy the PDW order and charge-2$e$ superconductivity. The composite uniform charge-4$e$ and charge-6$e$ order parameters [13], decoupled from the proliferating dislocations, can therefore become the primary quasi-long-range vestigial ordered states. The predicted charge-4$e$ state has chiral d+id symmetry and requires residual orientation order, whereas the charge-6$e$ state is isotropic with s-wave symmetry [13].

Intriguingly, the melting of the vortex-antivortex lattice into an isotropic liquid phase can indeed be staged and exhibit an intermediate liquid crystal phase, termed as a hexatic, with orientation order [40-42]. This happens when the topological defects associated with rotation, i.e. the disclinations, remain bound succeeding the proliferation of dislocations that restores translation symmetry. Our observation of charge-4*e* flux quantization in the intermediate temperature range on the small ring device [Fig. 2(c)] is thus consistent with such a hextic charge-4*e* state with the superconducting correlation length exceeding the circumference of the inner hole. Further increasing the temperature causes the unbinding and proliferation of the disclinations, which destroying the orientation order and the charge-4*e* state, as the melting into the isotropic liquid is completed, giving rise to the isotropic s-wave charge-6*e* superconductor. The sharp FFT peaks located at $6e/h$ for charge-6*e* flux quantization in Fig. 2(i) with the nearly perfect periodicity down to zero magnetic field are indeed consistent with the most robust s-wave charge-*6e* state in the isotropic phase proposed in this scenario. The chiral phase in the vestigial hexatic charge-4*e* state, on the other hand, couples strongly to the strain fields, supercurrent fluctuations, and disclination defects [13], which can limit the correlation length and hinders the ability of the charge-4*e* bound states to move through the ring structure phase coherently. As a result, in contrast to the robust isotropic charge-*6e* state, the charge-4*e* state is less robust, which is consistent with our observation that the $h/4e$ oscillations are relatively weak in small ring devices and difficult to discern in the micron-sized ring devices with much larger hole areas (Table 1). Moreover, a direct melting transition from the vortex-antivortex lattice to the isotropic liquid phase is also possible in the micron-sized ring devices, resulting in the single higher-charge vestigial ordered charge-6*e* superconductivity. Although the 3Q chiral PDW offers a physically intuitive picture for our observations, the origin of the observed charge-6*e* flux quantization is wide open for future investigations, as is the mechanism for superconductivity and PDW order in the kagome superconductors.

In summary, we discovered quantum oscillations with the periodicity of multi-charge flux quantization in mesoscopic $CsV_3Sb_5$ superconducting ring devices, which suggests the possibility of higher-charge superconductivity. Our experimental findings bring new insights into the rich and fascinating quantum states in the kagome superconductors and provide ground work for exploring the physical properties of unprecedented phases of matter formed by multi-particle bound states.

## ACKNOWLEDGMENTS

A portion of this work was carried out at the Synergetic Extreme Condition User Facility (SECUF). We acknowledge discussions with Li Lu, Yanzhao Liu, Yi Liu, Yanan Li, Haoran Ji and Jingchao Fang, and technical assistance from Jiawei Luo, Pengfei Zhan, Chunsheng Gong, Zhijun Tu, Yinbo Ma, and Gaoxing Ma. This work was financially supported by the National Natural Science Foundation of China (Grant No. 12488201), the National Key Research and Development Program of China (Grant No. 2018YFA0305604, No. 2018YFE0202600), the National Natural Science Foundation of China (Grant No. 11974430), Beijing Natural Science Foundation (Z180010, Z200005), the Innovation Program for Quantum Science and Technology (2021ZD0302403) and the China Postdoctoral Science Foundation (2022M720270). Z.Q.W. acknowledges support from the U.S. Department of Energy, Basic Energy Sciences Grant No. DE-FG02-99ER45747 and the Cottrell SEED Award No. 27856 from Research Corporation for Science Advancement.

## APPENDIX A: MATERIALS AND METHODS

**Crystal growth.** Single crystals of $CsV_3Sb_5$ were grown by using the self-flux method [28]. In a typical growth, the mixture of Cs ingot (purity 99.75%), V powder (purity 99.9%) and Sb grains (purity 99.999%) was put into an alumina crucible and sealed in a quartz ampoule under argon atmosphere. The quartz ampoule was heated up to 1273 K for 12 h and held for 24 h. Then it was rapidly cooled down to 1173 K in two hours and slowly cooled down to 923 K. Finally, the $CsV_3Sb_5$ single crystals were separated from the flux by using a centrifuge. In order to prevent the reaction of Cs with air and water, all the preparation processes except the sealing and heat treatment procedures were carried out in an argon-filled glove box.

**Devices fabrication.** The thin $CsV_3Sb_5$ flakes were firstly exfoliated from bulk single crystals using the scotch tape and then transferred onto the 300-nm-thick $SiO_2$/Si substrates, which were pre-cleaned in oxygen plasma for 5 min at ~70 mTorr pressure. By a standard electron beam lithography process in a FEI Helios NanoLab 600i Dual Beam System with PMMA 495A11 as a resist, electrodes were patterned, and metal electrodes (Ti/Au, 6.5/180 nm) were deposited in a LJUHVE-400 L E-Beam Evaporator. Note that the PMMA 495A11 resist was not baked. Finally, the PMMA 495A11 layers were removed by standard lift-off process and $CsV_3Sb_5$ flake devices with four metal electrodes were obtained. All the device fabrication processes were carried out in an argon-filled glove box with the $O_2$ and $H_2O$ levels below 0.1 ppm.

Then, PMMA 495A11 electron-beam resist (800 nm thick, without baking) was spin coated on the $CsV_3Sb_5$ flake devices as the protection layers for etching the $CsV_3Sb_5$ ring structures by focused ion beam (FIB) technique. The beam current for etching was 7.7 pA. FIB etching was carried out in FEI Helios NanoLab 600i Dual Beam System.

The Nb thin films with thickness of ~ 30 nm were deposited by magnetron sputtering. Then, PMMA 495A11 electron-beam resist (800 nm thick, without baking) was spin-coated as the protection layer for etching ring structures on the Nb thin films by FIB in a FEI Helios NanoLab 600i Dual Beam System. The beam current for etching was 7.7 pA.

**Transport measurements.** The transport measurements for $CsV_3Sb_5$ ring devices s1 were conducted with the dilution refrigerator (DR) option with working temperature from 4 K to 0.05 K, and the transport measurements for $CsV_3Sb_5$ ring device s2-s4 were conducted with $^3$He cryostat option in a 16 T-Physical Property Measurement System (PPMS-16, Quantum Design). The transport measurements for $CsV_3Sb_5$ ring device s5 were conducted with a Oxford instruments dilution refrigerator with temperature range ~10 mK – 30 K. The transport measurements of Nb ring devices were carried out with the Resistivity option in a 16 T-Physical Property Measurement System (PPMS-16, Quantum Design). The DC excitation current is 0.5 μA for $CsV_3Sb_5$ ring device s1 and s3, 2 μA for $CsV_3Sb_5$ ring device s2, 1 μA for $CsV_3Sb_5$ ring device s4 and s5, 2 μA for Nb ring devices n1 and n2.

**AFM measurements.** The AFM measurements of the $CsV_3Sb_5$ devices were performed in the Bruker DimensionICON system.

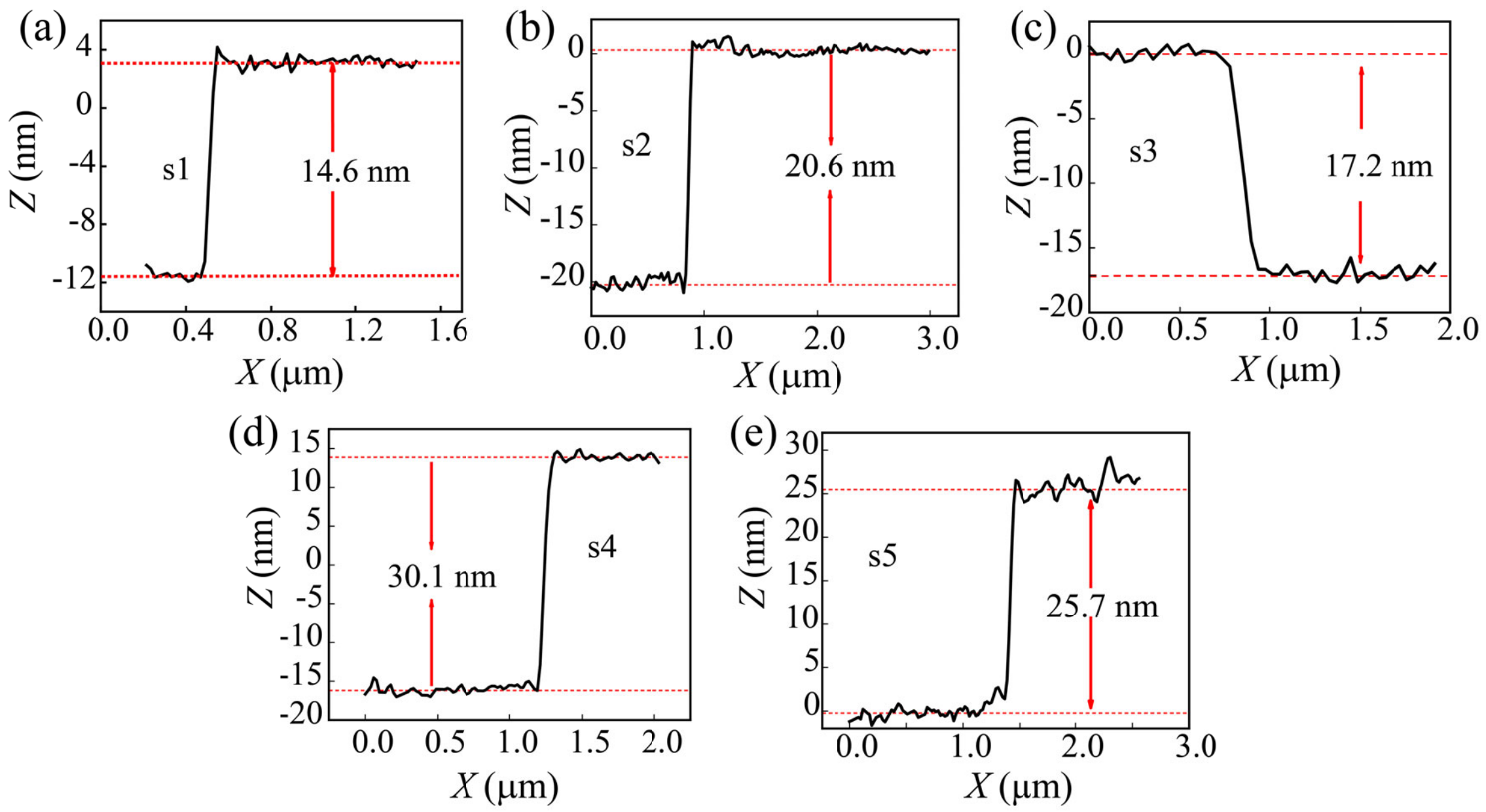

FIG. 5. Thickness of $CsV_3Sb_5$ ring device s1 (a), s2 (b), s3 (c), s4 (d) and s5 (e) measured by atomic force microscopy.

**Subtracting magnetoresistance backgrounds.** The magnetoresistance backgrounds are obtained by properly smoothing the raw *R*-*H* curves. We take the raw *R*-*H* curve of s1 at 2.4 K with $h/6e$ oscillations as an example to show the detailed data process procedure. As shown in Fig. 6(a), the background is obtained by 80-points Savitzky-Golay smoothing of the raw data (shown by red line). The oscillation signals (inset of Fig. 6(a)) are eventually obtained by subtracting the red line from the raw data.

We have also drawn the n-$H_n$ index plot from the raw *R*-*H* curve, as shown in Fig. 6(b). By linearly fitting the n-$H_n$ plot, a period of ~251 Oe is obtained, which is consistent with the period obtained from the oscillation signals after subtracting the background (approximately 250 Oe). Different methods have revealed consistent period, further verifying that the determination of the magnetoresistance background is reliable.

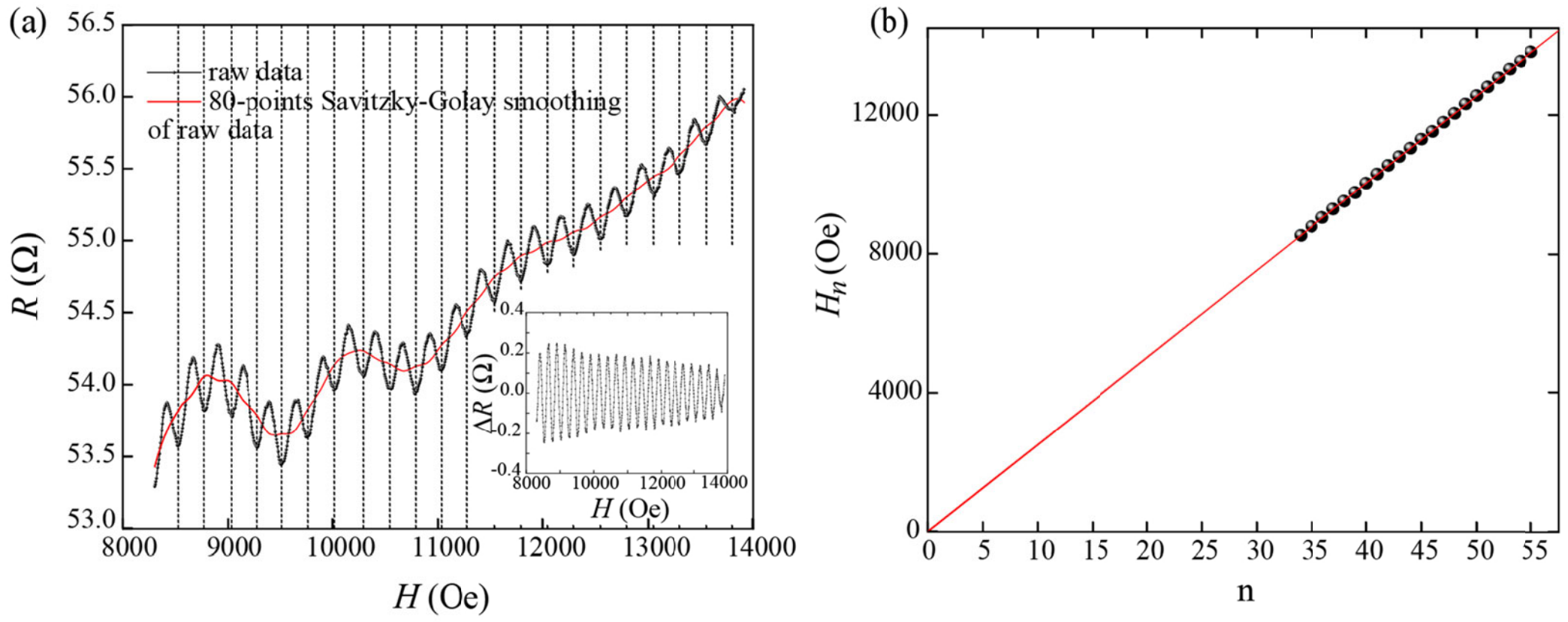

FIG. 6. (a) Raw resistance (*R*)-magnetic field (*H*) curve showing obvious $h/6e$ oscillations in $CsV_3Sb_5$ ring device s1. The background is obtained by 80-points Savizkg-Golag smoothing of the raw data (shown by red line). The oscillation signals (inset of (a)) are eventually obtained by subtracting the red line from the raw data. (b) n-$H_n$ index plot obtained from the dashed lines in (a). The linear fitting reveals a period of ~251 Oe for the oscillations, which is consistent with the period obtained from the oscillation signals after subtracting the background (approximately 250 Oe), indicating that the method of subtracting the background is reliable.

## APPENDIX B: THEORETICAL DISCUSSIONS ON FRACTIONAL FLUX QUANTUM IN MULTICOMPONENT SUPERCONDUCTORS

It was theoretically proposed that multicomponent superconductors can host fractional flux quantum [39]. The multicomponents can come from different bands/Fermi surfaces, spins, or even coupled with different momenta as in the case of pair density wave superconductors. The basic idea is that in a magnetic field that causes the phase winding of the multicomponent order parameters, if one or more components do not wind, or wind differently, then the single-valuedness of the wave function results in a fractionally quantized flux quantum. This seemingly simple proposal has not been realized over the years, possibly because that in a real superconductor, these multicomponent condensates are usually phase-locked together by the Josephson coupling, and the phases of different order parameter components wind together to give the expected flux quantization.

Consider the p+ip superconductors as an example. This is a spin-triplet pairing state with equal-spin pairing. The order parameter is a mixing of |1,1> and |2,2>, with 1 and 2 denoting the two spin components. The single-valuedness of the wave function requires the flux quantization $\Phi = \Phi_0 (n_1+n_2)/2$, where $\Phi_0 = h/2e$ is the usual flux quantum, and $n_1$ and $n_2$ are the integer number of times that each spin-component winds around $2\pi$. Consider the case where $n_1$ and $n_2$ are both nonzero, but differ by an odd integer due to the different Berry phases of electrons with opposite spins, such that $n_1 + n_2 = 2n + 1$ and $\Phi = \Phi_0 (n+1/2)$, giving rise to the half-flux-quantum (HFQ). Note that this HFQ only causes a shift in $\Phi$, leading to a $\pi$ phase shift of the quantum oscillations, while the period of oscillations would remain to be $\Phi_0$. This is, however, very different from our observations of the quantum oscillations with $h/6e$ periodicity corresponding to $\frac{\Phi 0}{3}$. Furthermore, the existing experimental evidence does not support spin-triplet pairing in kagome superconductors [27,32,43-45].

Next consider the case where one component doesn't wind, i.e. either $n_1$ or $n_2$ equals zero. Then, $\Phi = \frac{\Phi 0}{2}$ times an integer. This can in principle lead to quantum oscillations with a flux quantization $h/4e$. This picture can be generalized to superconductors with N number of degenerate components. Depending on the various hypothesized inter-component Josephson couplings, oscillations with fractional flux quantization is possible and mostly favored by $\Phi_0/N$. Our observations do not rule out such a scenario, but the observation of the temperature evolution of the flux quantization from $h/2e$ to $h/6e$ makes it unnatural even at a phenomenological level. We note that in all the scenarios discussed above, the pairing is still charge-$2e$ in nature, i.e. a Cooper pair, whereas our observations suggest more naturally that the successive transitions in the periodicity of the quantum oscillations is connected to charge-$6e$ pairing. Furthermore, to our knowledge, oscillations with $h/4e$ periodicity have never been experimentally reported in p+ip system.

Hexagonal pair density waves (PDWs) turn out to offer a more physically intuitive account of our observations, in addition to having been observed in bulk $CsV_3Sb_5$ [23]. A PDW is also a multicomponent superconductor. The difference is that each component in a PDW also carries a well-defined momentum. This opens the possibility for charge-2$e$ PDW induced vestigial higher-charge off-diagonal long-range order [6,7,9].

## APPENDIX C: ROBOUST $h/6e$ OSCILLATIONS IN $CsV_3Sb_5$ RING DEVICES

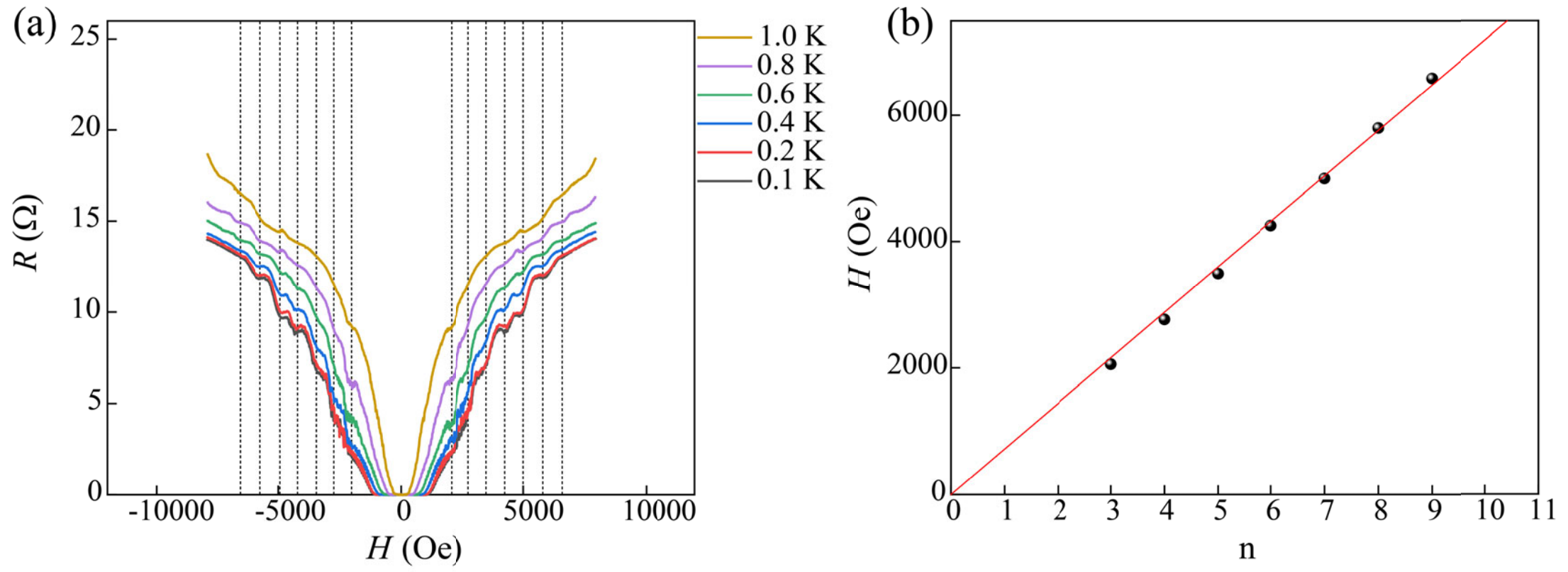

FIG. 7. $h/2e$ oscillations in $CsV_3Sb_5$ ring device s1. (a) Raw $R$-$H$ curves at various temperatures. The dashed lines label the dips of the oscillations and serve as guides to the eye. (b) n-$H_n$ index plot of the $h/2e$ oscillations. Here, n represents the $n_{th}$ oscillation dip. A linear relation between n and $H_n$ can be clearly observed, suggesting the oscillations are periodic.

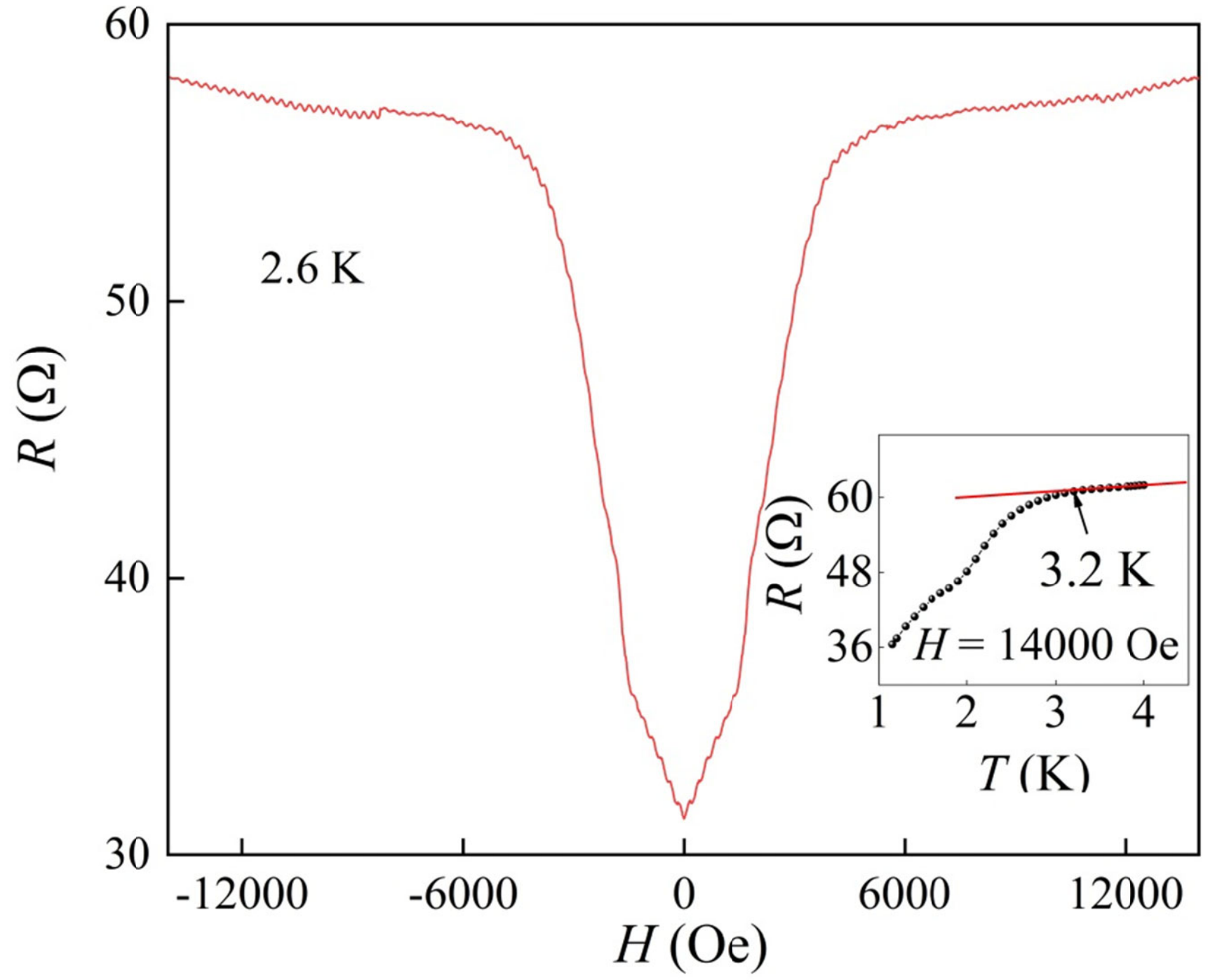

FIG. 8. Oscillations with periodicity of $h/6e$ in $CsV_3Sb_5$ ring device s1 in a wide field regime from -14000 Oe to 14000 Oe. Inset shows the $R$-$T$ curve at 14000 Oe. The superconductivity drop with $T_c^{onset}$ of ~3.2 K can be detected in the $R$-$T$ curve. The $h/6e$ oscillations appearing below 1.4 T and 3.0 K are in the superconducting transition region.

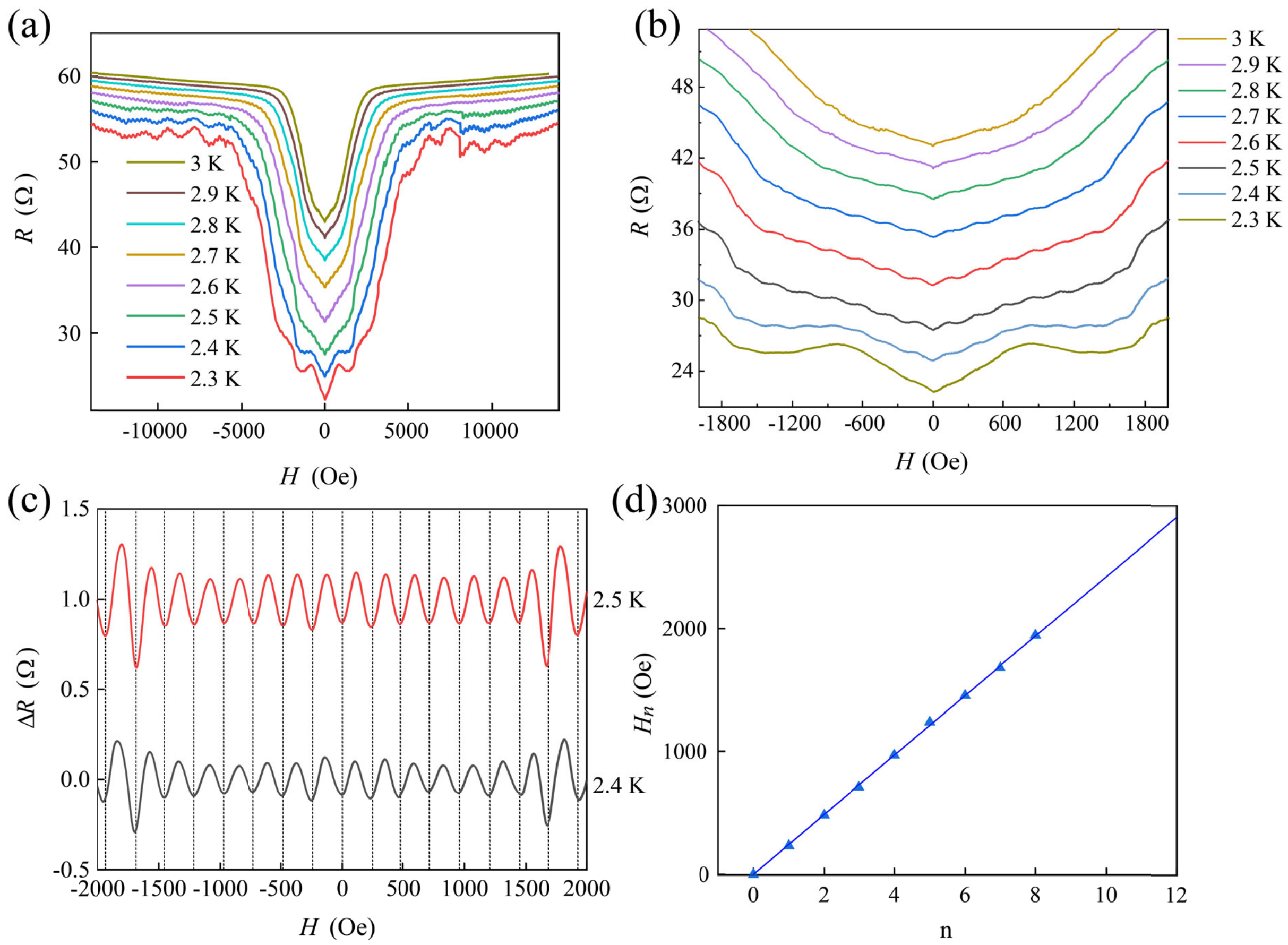

FIG. 9. Oscillations with periodicity of $h/6e$ in $CsV_3Sb_5$ ring device s1 at various temperatures. (a) Raw *R*-*H* curves at various temperatures. The $h/6e$ oscillations (small period) superimposed on another set of large-period oscillations (details shown in Fig. 10) appear in the superconducting transition region and vanish at 3 K. (b) *R*-*H* curves near zero magnetic field at various temperatures. (c) Resistance oscillations with $h/6e$ periodicity in which the rising background has been subtracted. The black dashed lines label the dips of the oscillations. (d) n-$H_n$ index plot of the $h/6e$ oscillations. A linear relation between n and $H_n$ can be clearly observed, indicating the good periodicity of the $h/6e$ oscillations.

## APPENDIX D: *h/e* OSCILLATIONS COEXISTING WITH *h*/6*e* OSCILLATIONS IN $CsV_3Sb_5$ RING DEVICE S1

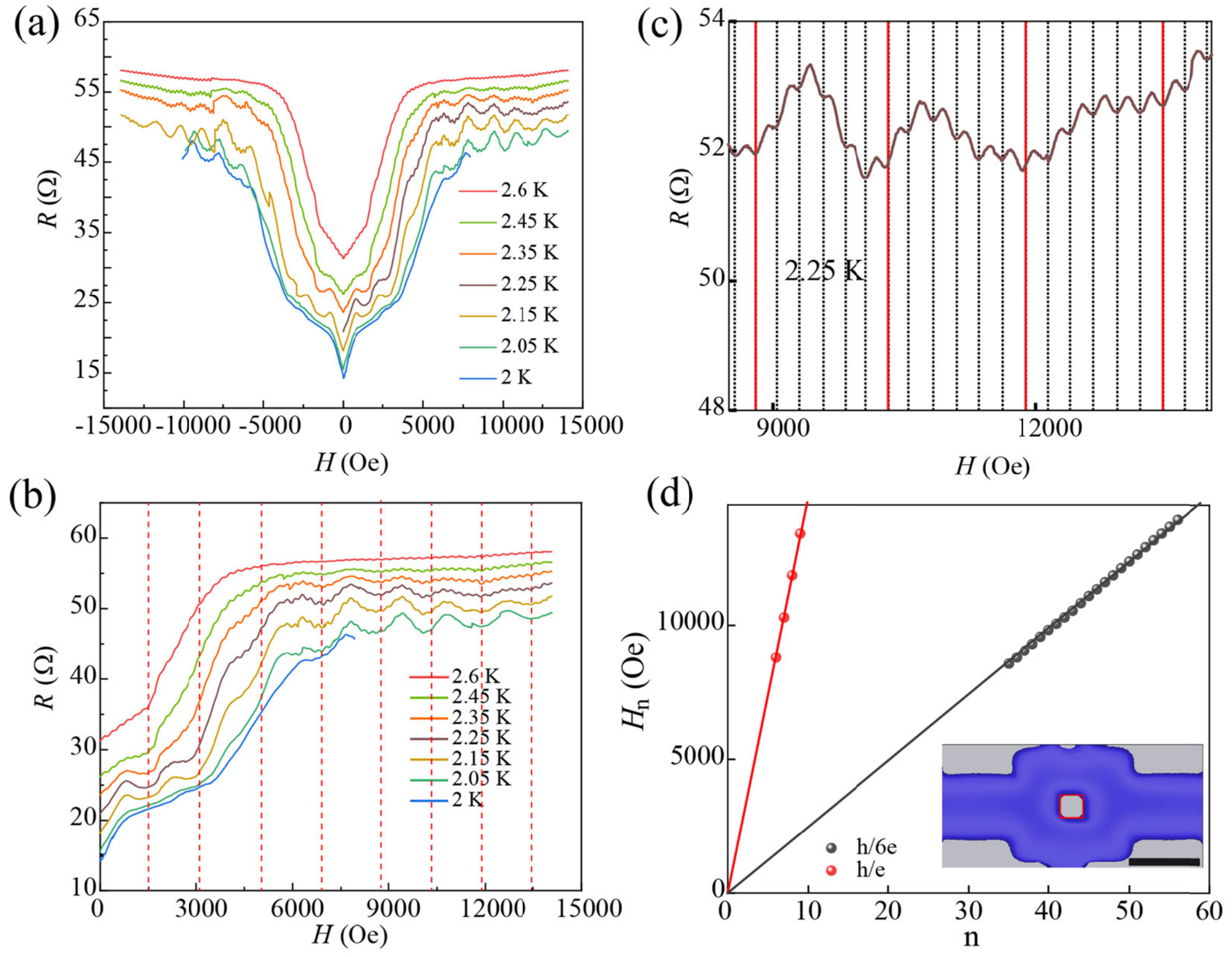

FIG. 10. Quantum oscillations with *h*/*e* and *h*/6*e* periodicity in s1. (a) Resistance (*R*) as a function of magnetic field (*H*) at temperatures from 2 K to 2.6 K. *h/6e* oscillations (small period) superimposed on another set of large-period oscillations can be observed on the raw *R*-*H* curves. All oscillations are observed in the superconducting transition region. (b) Raw *R*-*H* curves at positive magnetic field. The red dashed lines label the oscillation dips of large-period oscillations. (c) Zoom-in of the raw *R*(*H*) curve at 2.25 K. The black dashed lines and red solid lines label the oscillation dips of *h*/6*e* oscillations (small period) and large-period oscillations, respectively. The period of the large-period oscillations is six times the period of the *h*/6*e* oscillations. Considering the same effective area ($S \approx 0.0276\ \mu m^2$) as the *h*/2*e*, *h*/4*e*, and *h*/6*e* oscillations in this ring device, the large-period oscillation (period ~1500 Oe) corresponds to a magnetic flux through the ring $\Phi = S \cdot \Delta H = 0.0276\ \mu m^2 \times 1500\ Oe = 4.14 \times 10^{-15}\ Wb = h/e$, yielding a periodicity with *h*/*e* flux quantization. (d) $H_n$-n index plot of the quantum oscillations in (c). Here, n represents the $n_{th}$ oscillation dip, and the corresponding magnetic field is identified as $H_n$. Inset shows the false-colored SEM image of s1. The sample protected by PMMA layer is represented by blue and the substrate is represented by gray. The scale bar in the false-colored SEM image represents 500 nm. The red rectangle marks the effective area for *h*/6*e* and *h*/*e* oscillations. Note

that if the effective area for the large-period *h*/*e* oscillation is larger (>0.0276 μm$^2$), the charge must be fractional, which is very unlikely to emerge in vanadium-based kagome superconductors such as $CsV_3Sb_5$. The *h*/*e* oscillations have long been theoretically predicted in ring structures of unconventional superconductors [46], wherein the inner hole area is identified as the effective area. Though our observation of the *h*/*e* oscillation with the effective area close to the inner hole area appears to be consistent with the theoretical predictions, the detailed origin of the *h*/*e* oscillations is not understood and warrants further investigation.

## APPENDIX E: COMPLETE TEMPERATURE EVOLUTION OF THE FLUX QUANTIZATION IN $CsV_3Sb_5$ RING DEVICES

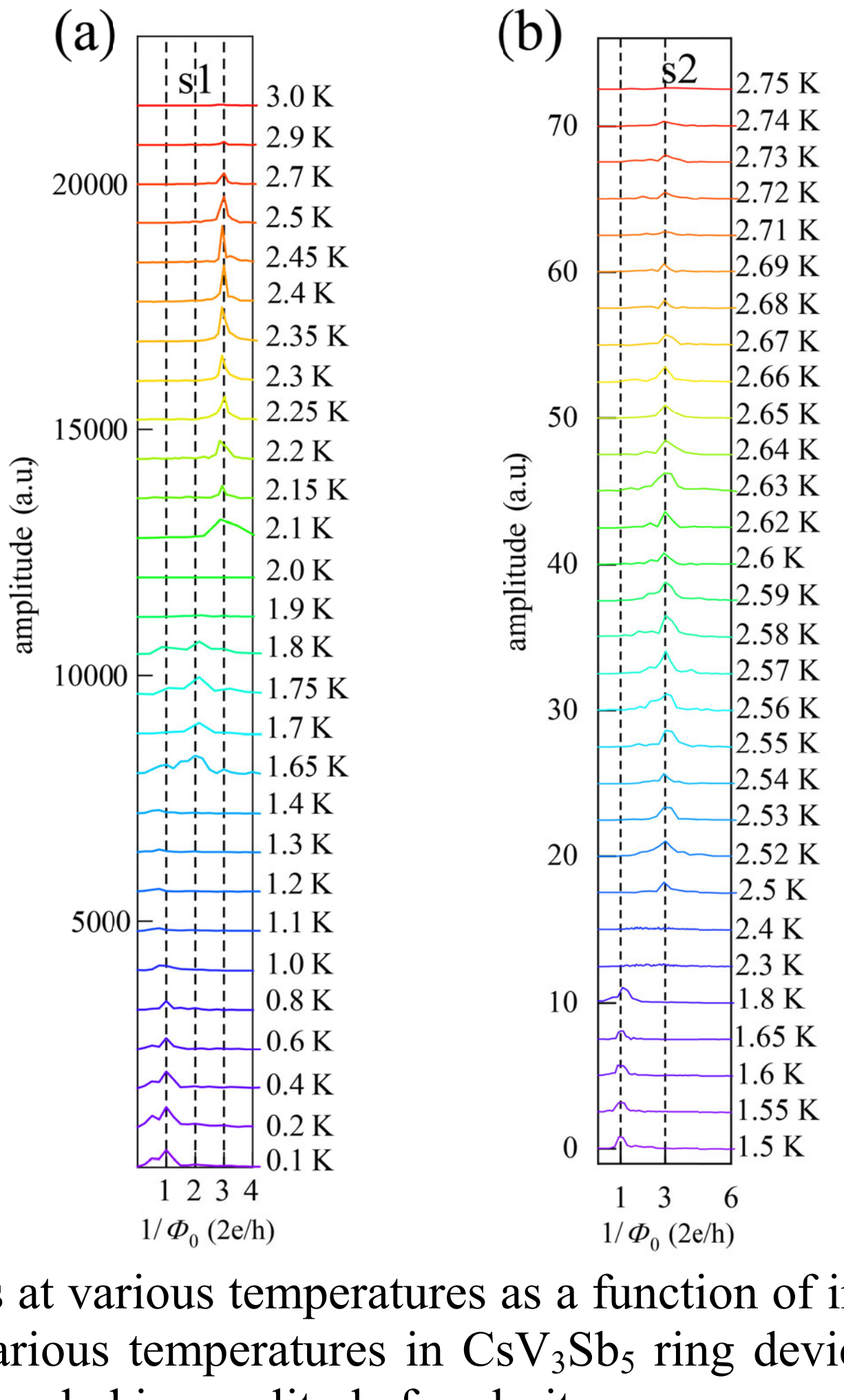

FIG. 11. FFT results at various temperatures as a function of inverse magnetic flux in unit of 2*e*/*h* at various temperatures in $CsV_3Sb_5$ ring device s1 (a) and s2 (b). The FFT curves are scaled in amplitude for clarity.

# APPENDIX F: MULTI-CHARGE FLUX QUANTITATION IN OTHER $CsV_3Sb_5$ RING DEVICES

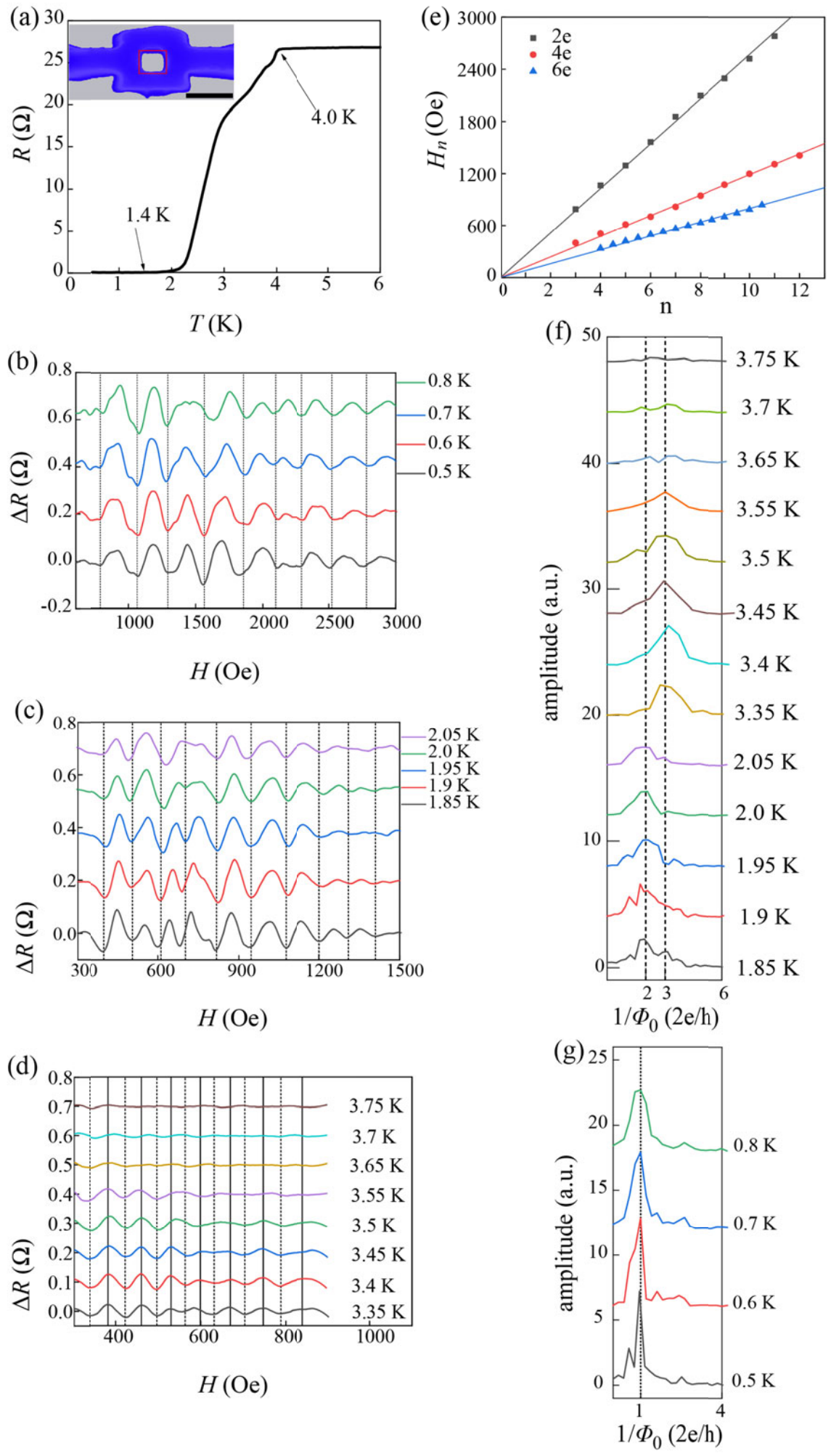

FIG. 12. Oscillations with periodicities of $h/2e$, $h/4e$ and $h/6e$ in superconducting $CsV_3Sb_5$ ring device s3. (a) Resistance as a function of temperature from 6 K to 0.5 K. Superconductivity with onset temperature ($T_c^{onset}$) ~4.0 K and zero-resistance temperature $T_c^{zero}$~1.4 K is observed. Inset shows the false-colored SEM image of s3. The sample protected by PMMA layer is represented by blue and the substrate is represented by gray. The scale bar in the false-colored SEM image represents 500 nm. The inner area, middle area and outer area of the ring estimated from the false-colored image is ~0.047 $\mu m^2$, ~0.21 $\mu m^2$ and ~0.57 $\mu m^2$, respectively. (b) $h/2e$ oscillations with period of ~ 255 Oe at different temperatures after subtracting smooth backgrounds. The dashed lines label the resistance dips. For clarity, data curves are shifted. The effective area of the $h/2e$ oscillations (~0.081 $\mu m^2$) is marked by the red rectangle in the inset of (a). (c), (d) $h/4e$ and $h/6e$ oscillations at different temperatures after subtracting smooth backgrounds,

respectively. The dashed and solid lines label the resistance dips and peaks, respectively. The period of the $h/4e$ and $h/6e$ oscillations is ~ 119 Oe and ~80 Oe, respectively. For clarity, data curves in (c) and (d) are shifted. (e) n-$H_n$ index plot of the $h/2e$, $h/4e$ and $h/6e$ oscillations. (f), (g) FFT results of the oscillations at various temperatures. Note that the oscillation range for FFT is not exactly the same as the range shown in (b)-(d). When the temperature is increased, the periodicity changes from $h/2e$ to $h/4e$ and then to $h/6e$.

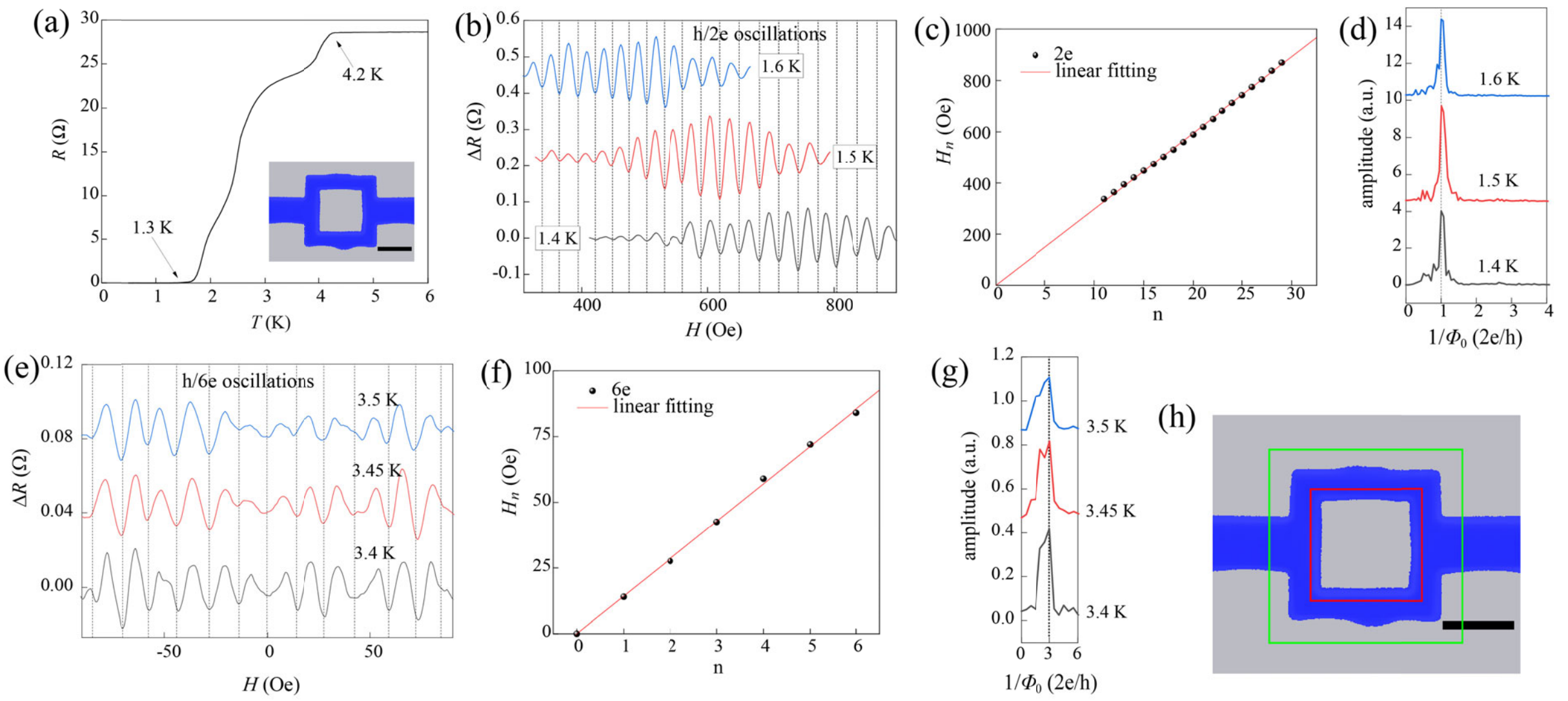

FIG. 13. *h*/2*e* and *h*/6*e* oscillations in the $CsV_3Sb_5$ ring device s4. (a) Resistance as a function of temperature from 6 K to 0.5 K. Superconductivity with onset temperature ~4.2 K and zero-resistance temperature ~1.3 K is observed. Inset shows the false-colored image of the $CsV_3Sb_5$ ring device s4. The sample protected by PMMA layer is represented by blue and the substrate is represented by gray. The scale bar in the false-colored image represents 500 nm. The inner area, middle area and outer area of the ring estimated from the false-colored SEM image is ~0.39 μm$^2$, ~ 0.77 μm$^2$ and ~ 1.15 μm$^2$, respectively. (b) *h*/2*e* oscillations with period of ~34 Oe after subtracting smooth backgrounds in the $CsV_3Sb_5$ ring device s4. The effective area of the *h*/2*e* oscillations is ~0.61 μm$^2$, which is marked by the red rectangle in (h). The dashed lines label the oscillation dips. For clarity, data curves are shifted. (c) n-$H_n$ index plot of the *h*/2*e* oscillations. Here, n is an integer, representing the oscillation dip. A linear relation between n and $H_n$ can be clearly observed, indicating that the magnetoresistance oscillations are periodic. (d) FFT results of the *h*/2*e* oscillations. Note that the oscillation range for FFT is not exactly the same as the range shown in (b). (e)-(g) Oscillations with period of ~11 Oe at higher temperatures. Note that the oscillation range for FFT in (g) is not exactly the same as the range shown in (e). The dashed lines in (e) label the oscillation dips. For clarity, data curves are shifted. The n-$H_n$ index plot shown in (f) indicates that the oscillations are periodic. If the oscillations with period of ~ 11 Oe were *h*/2*e* oscillations, then the effective area (1.88 μm$^2$, shown by the green rectangle in (h)) would be obviously much larger than the outer area of the ring device (~1.15 μm$^2$), which is physically impossible. Considering the same effective area as the *h*/2*e* oscillations with period of ~34 Oe, the oscillations with period of ~11 Oe corresponds to a *h*/6*e* periodicity.

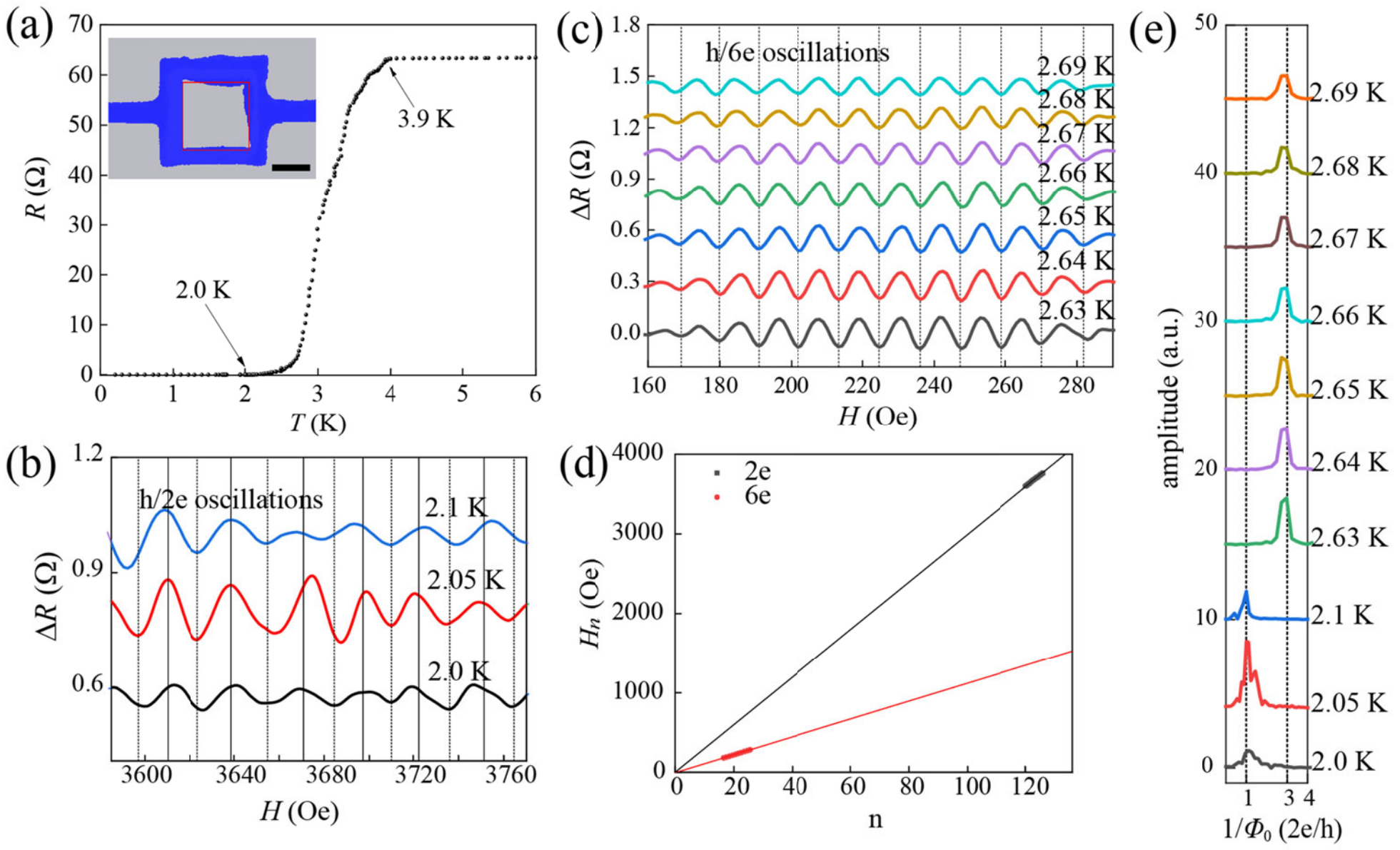

FIG. 14. $h/2e$ and $h/6e$ oscillations in the $CsV_3Sb_5$ ring device s5. (a) Resistance as a function of temperature from 6 K to 0.2 K. Superconductivity with onset temperature ~ 3.9 K and zero-resistance temperature ~2.0 K is observed. Inset shows the false-colored image of the $CsV_3Sb_5$ ring device s5. The sample protected by PMMA layer is represented by blue and the substrate is represented by gray. The scale bar in the false-colored image represents 500 nm. The inner area, middle area and outer area of the ring estimated from the false-colored SEM image is ~0.64 μm$^2$, ~1.20 μm$^2$ and ~1.78 μm$^2$, respectively. (b) $h/2e$ oscillations with period of ~30 Oe after subtracting smooth backgrounds in the $CsV_3Sb_5$ ring device s5. The effective area of the $h/2e$ oscillations is ~0.69 μm$^2$, which is marked by the red rectangle in the inset of (a). The dashed and solid lines label the oscillation dips and peaks, respectively. For clarity, data curves are shifted. (c) Oscillations with period of ~11 Oe at higher temperatures after subtracting smooth backgrounds. The dashed lines label the oscillation dips. For clarity, data curves are shifted. If the oscillations with period of ~11 Oe were $h/2e$ oscillations, the effective area would be ~ 1.88 μm$^2$, larger than the outer area of the ring device (~1.78 μm$^2$), which is physically impossible. Considering the same effective area as the $h/2e$ oscillations with period of ~30 Oe, the oscillations with period of ~11 Oe corresponds to a $h/6e$ periodicity. (d) n-$H_n$ index plot of the $h/2e$ and $h/6e$ oscillations. Here, n is an integer or half integer, representing the oscillation dip and peak, respectively. The linear relation between n and $H_n$ confirms that the magnetoresistance oscillations are periodic. (e) FFT results of the $h/2e$ and $h/6e$ oscillations. Note that the oscillation range for FFT is not exactly the same as the range shown in (b), (c).

TABLE 1. An overview of five $CsV_3Sb_5$ ring devices with different sizes and the flux quantizations observed in these ring devices. The scale bar in each false-colored SEM image represents 500 nm and the red rectangle marks the effective area consistent with all the flux quantizations of the magnetoresistance oscillations observed in the corresponding ring device.

| Sample number | SEM image | Thickness (nm) | Inner hole area ($\mu m^2$) | Outer ring area ($\mu m^2$) | Effective area ($\mu m^2$) | Flux quantizations |
|---|---|---|---|---|---|---|
| s1 | 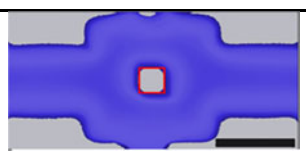 | 14.6 | 0.027 | 0.54 | 0.0276 | *h*/2*e*, *h*/4*e*, *h*/6*e* |
| s2 | 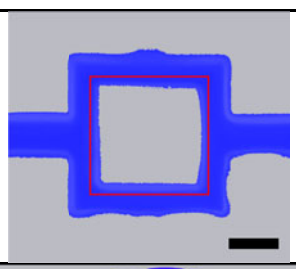 | 20.6 | 0.96 | 2.97 | 1.52 | *h*/2*e*, *h*/6*e* |
| s3 | 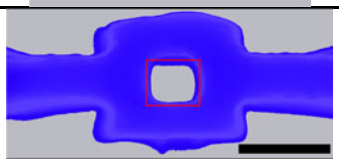 | 17.2 | 0.047 | 0.57 | 0.081 | *h*/2*e*, *h*/4*e*, *h*/6*e* |
| s4 | 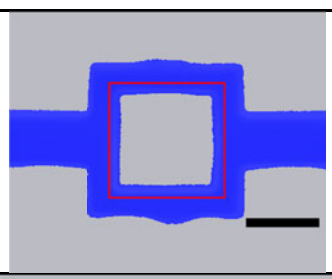 | 30.1 | 0.39 | 1.15 | 0.61 | *h*/2*e*, *h*/6*e* |
| s5 | 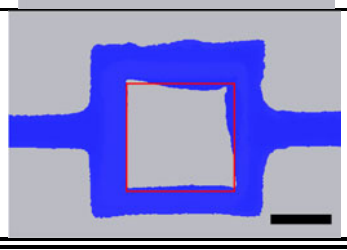 | 25.7 | 0.64 | 1.78 | 0.69 | *h*/2*e*, *h*/6*e* |

## APPENDIX G: CONTROL EXPERIMENT RESULTS OF CONVENTIONAL SUPERCONDUCTOR Nb RING DEVICES ETCHED BY SAME METHOD AS $CsV_3Sb_5$

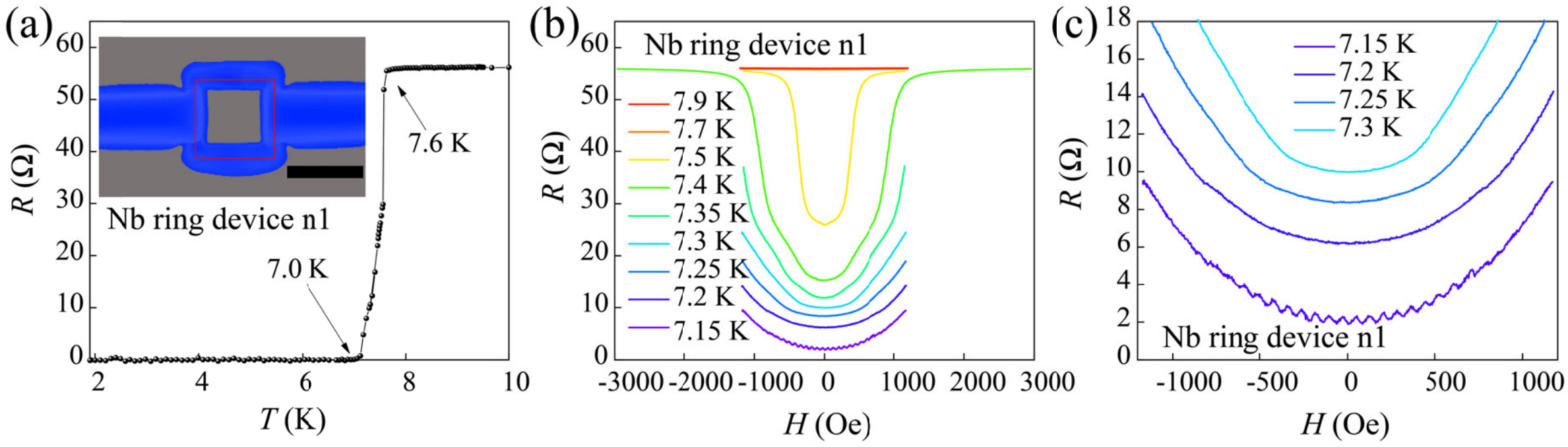

FIG. 15. *h*/2*e* oscillations in conventional superconductor Nb ring device n1 etched by FIB. (a) The resistance (*R*)-temperature (*T*) curve of the Nb ring device n1. A sharp superconducting transition with onset temperature of ~7.6 K and zero

resistance temperature of ~7.0 K is observed. Inset shows the false-colored SEM image of the Nb ring device n1. The sample protected by PMMA layer is represented by blue and the substrate is represented by gray. The scale bar represents 500 nm. The inner area of the ring estimated from the false-colored SEM image is ~0.120 μm$^2$. (b) Magnetoresistance of the Nb ring device. Only $h/2e$ oscillations in a narrow temperature regime (around 7.15 K) were observed in the Nb ring device. The effective area of the $h/2e$ oscillations is ~0.258 μm$^2$, which is marked by the red rectangle in the inset of (a). (c) $h/2e$ oscillations in the Nb ring device.

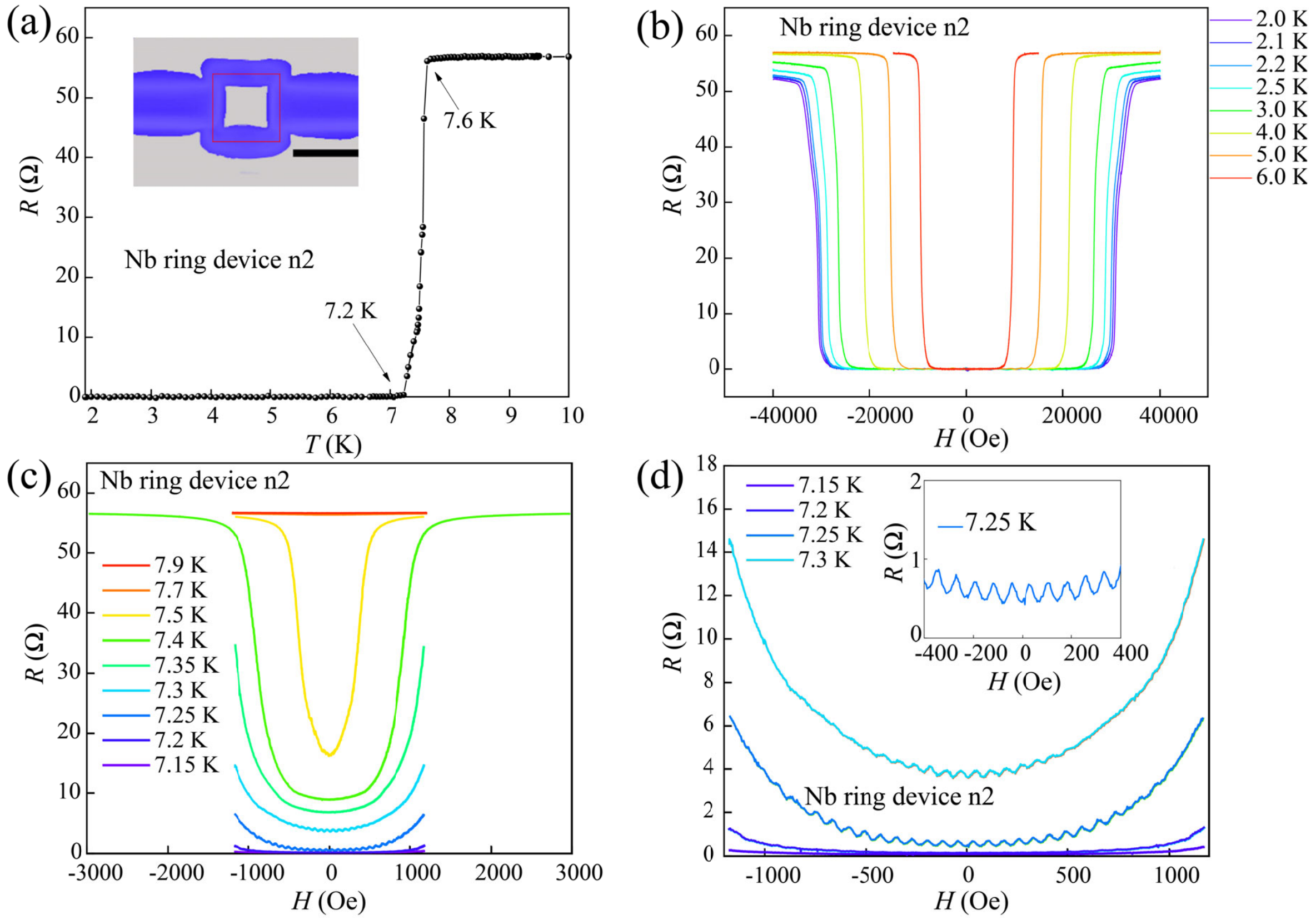

FIG. 16. $h/2e$ oscillations in conventional superconductor Nb ring device n2 etched by FIB. (a) The resistance ($R$)-temperature ($T$) curve of the Nb ring device. A sharp superconducting transition with onset temperature of ~7.6 K and zero resistance temperature of ~7.2 K is observed. Inset shows the false-colored SEM image of the Nb ring device n2. The sample protected by PMMA layer is represented by blue and the substrate is represented by gray. The scale bar represents 500 nm. The inner area of the ring estimated from the false-colored SEM image is ~0.106 μm$^2$. (b), (c) Magnetoresistance of the Nb ring device at low (b) and high temperatures (c). Systematic and careful measurements on the magnetoresistance at different temperatures were carried out. Only $h/2e$ oscillations in a narrow temperature

regime (from 7.25 to 7.3 K) were observed in the Nb ring device. The effective area of the *h*/2*e* oscillations is ~0.268 μm$^2$, which is marked by the red rectangle in the inset of (a). (d) *h*/2*e* oscillations in the Nb ring device.

[1] J. R. Schrieffer, *Theory of Superconductivity Perseus Books*, Reading (1999).

[2] J. Bardeen, L. N. Cooper, and J. R. Schrieffer, *Theory of Superconductivity*, Phys. Rev. **108**, 1175 (1957).

[3] B. S. Deaver and W. M. Fairbank, *Experimental Evidence for Quantized Flux in Superconducting Cylinders*, Phys. Rev. Lett. **7**, 43 (1961).

[4] R. Doll and M. Näbauer, *Experimental Proof of Magnetic Flux Quantization in a Superconducting Ring*, Phys. Rev. Lett. **7**, 51 (1961).

[5] N. Byers and C. N. Yang, *Theoretical Considerations Concerning Quantized Magnetic Flux in Superconducting Cylinders*, Phys. Rev. Lett. **7**, 46 (1961).

[6] D. F. Agterberg and H. Tsunetsugu, *Dislocations and Vortices in Pair-Density-Wave Superconductors*, Nat. Phys. **4**, 639 (2008).

[7] E. Berg, E. Fradkin, and S. A. Kivelson, *Charge 4e Superconductivity from Pair Density Wave Order in Certain High Temperature Superconductors*, Nat. Phys. **5**, 830 (2009).

[8] L. Radzihovsky and A. Vishwanath, *Quantum Liquid Crystals in an Imbalanced Fermi Gas: Fluctuations and Fractional Vortices in Larkin-Ovchinnikov States*, Phys. Rev. Lett. **103**, 010404 (2009).

[9] E. V. Herland, E. Babaev, and A. Sudbø, *Phase Transitions in a Three Dimensional U(1)×U(1) Lattice London Superconductor: Metallic Superfluid and Charge-4e Superconducting States*, Phys. Rev. B **82**, 134511 (2010).

[10] D. F. Agterberg, M. Geracie, and H. Tsunetsugu, *Conventional and charge-six superfluids from melting hexagonal Fulde-Ferrell-Larkin-Ovchinnikov phases in two dimensions*, Phys. Rev. B **84**, 014513 (2011).

[11] S.-K. Jian, Y. Huang, and H. Yao, *Charge-4e Superconductivity from Nematic Superconductors in Two and Three Dimensions*, Phys. Rev. Lett. **127**, 227001 (2021).

[12] R. M. Fernandes and L. Fu, *Charge-4e Superconductivity from Multicomponent Nematic Pairing: Application to Twisted Bilayer Graphene*, Phys. Rev. Lett. **127**, 047001 (2021).

[13] S. Zhou and Z. Wang, *Chern Fermi Pocket, Topological Pair Density Wave, and Charge-4e and Charge-6e Superconductivity in Kagome Superconductors*, Nat. Commun. **13**, 7288 (2022).

[14] E. Fradkin, S. A. Kivelson, and J. M. Tranquada, *Colloquium: Theory of Intertwined Orders in High Temperature Superconductors*, Rev. Mod. Phys. **87**, 457 (2015).

[15] D. F. Agterberg, J. S. Davis, S. D. Edkins, E. Fradkin, D. J. V. Harlingen, S. A. Kivelson, P. A. Lee, L. Radzihovsky, J. M. Tranquada, and Y. Wang, *The Physics of Pair-Density Waves: Cuprate Superconductors and beyond*, Annu. Rev. Condens. Matter Phys. **11**, 231 (2020).

[16] R. M. Fernandes, P. P. Orth, and J. Schmalian, *Intertwined Vestigial Order in Quantum Materials: Nematicity and beyond*, Annu. Rev. Condens. Matter Phys. **10**, 133 (2019).

[17] B. R. Ortiz, L. C. Gomes, J. R. Morey, M. Winiarski, M. Bordelon, J. S. Mangum, I. W. H. Oswald, J. A. Rodriguez-Rivera, J. R. Neilson, S. D. Wilson, E. Ertekin, T. M. McQueen, and E. S. Toberer, *New Kagome Prototype Materials: Discovery of* $KV_3Sb_5$*,* $RbV_3Sb_5$*, and* $CsV_3Sb_5$, Phys. Rev. Mater. **3**, 094407 (2019).

[18] B. R. Ortiz, S. M. L. Teicher, Y. Hu, J. L. Zuo, P. M. Sarte, E. C. Schueller, A. M. M. Abeykoon, M. J. Krogstad, S. Rosenkranz, R. Osborn, R. Seshadri, L. Balents, J. He, and S. D. Wilson, $CsV_3Sb_5$*: A Z2 Topological Kagome Metal with a Superconducting Ground State*, Phys. Rev. Lett. **125**, 247002 (2020).

[19] C. C. Zhao, L. S. Wang, W. Xia, Q. W. Yin, J. M. Ni, Y. Y. Huang, C. P. Tu, Z. C. Tao, Z. J. Tu, C. S. Gong, H. C. Lei, Y. F. Guo, X. F. Yang, and S. Y. Li, *Nodal Superconductivity and Superconducting Domes in The Topological Kagome Metal* $CsV_3Sb_5$, arXiv:2102.08356.

[20] H. Zhao, H. Li, B. R. Ortiz, S. M. L. Teicher, T. Park, M. Ye, Z. Wang, L. Balents, S. D. Wilson, and I. Zeljkovic, *Cascade of Correlated Electron States in the Kagome Superconductor* $CsV_3Sb_5$, Nature (London) **599**, 216 (2021).

[21] F. Yu, T. Wu, Z. Wang, B. Lei, W. Zhuo, J. Ying, and X. Chen, *Concurrence of Anomalous Hall Effect and Charge Density Wave in a Superconducting Topological Kagome Metal*, Phys. Rev. B **104**, 041103 (2021).

[22] H. Tan, Y. Liu, Z. Wang, and B. Yan, *Charge Density Waves and Electronic Properties of Superconducting Kagome Metals*, Phys. Rev. Lett. **127**, 046401 (2021).

[23] H. Chen, H. Yang, B. Hu, Z. Zhao, J. Yuan, Y. Xing, G. Qian, Z. Huang, G. Li, and Y. Ye et al., *Roton Pair Density Wave in a Strong-Coupling Kagome Superconductor*, Nature (London) **599**, 222 (2021).

[24] Y.-X. Jiang, J.-X. Yin, M. M. Denner, N. Shumiya, B. R. Ortiz, J. He, X. Liu, S. S. Zhang, G. Chang, and I. Belopolski et al., *Discovery of Topological Charge Order in Kagome Superconductor* $KV_3Sb_5$, Nat. Mater. **20**, 1353 (2021).

[25] Z. Liang, X. Hou, W. Ma, F. Zhang, P. Wu, Z. Zhang, F. Yu, J.-J. Ying, K. Jiang, and L. Shan et al., *Three-Dimensional Charge Density Wave and Robust Zero-Bias Conductance Peak inside the Superconducting Vortex Core of a Kagome Superconductor* $CsV_3Sb_5$, Phys. Rev. X **11**, 031026 (2021).

[26] S.-Y. Yang, Y. Wang, B. R. Ortiz, D. Liu, J. Gayles, E. Derunova, R. Gonzalez-Hernandez, L. Smejkal, Y. Chen, and S. S. Parkin et al., *Giant,*

*Unconventional Anomalous Hall Effect in the Metallic Frustrated Magnet Candidate, $KV_3Sb_5$*, Sci. Adv. **6**, eabb6003 (2020).
[27] H.-S. Xu, Y.-J. Yan, R. Yin, W. Xia, S. Fang, Z. Chen, Y. Li, W. Yang, Y. Guo, and D.-L. Feng, *Multiband Superconductivity with Sign-Preserving Order Parameter in Kagome Superconductor $CsV_3Sb_5$*, Phys. Rev. Lett. **127**, 187004 (2021).
[28] K. Chen, N. Wang, Q. Yin, Z. Tu, C. Gong, J. Sun, H. Lei, Y. Uwatoko, and J.-G. Cheng, *Double Superconducting Dome and Triple Enhancement of Tc in the Kagome Superconductor $CsV_3Sb_5$ under High Pressure*, Phys. Rev. Lett. **126**, 247001 (2021).
[29] Y. Xiang, Q. Li, Y. Li, W. Xie, H. Yang, Z. Wang, Y. Yao, and H. H. Wen, *Twofold Symmetry of c-Axis Resistivity in Topological Kagome Superconductor $CsV_3Sb_5$ with In-Plane Rotating Magnetic Field*, Nat. Commun. **12**, 6727 (2021).
[30] C. Mielke, D. Das, J. X. Yin, H. Liu, R. Gupta, Y. X. Jiang, M. Medarde, X. Wu, H. C. Lei, and J. Chang et al., *Time-reversal symmetry-breaking charge order in a kagome superconductor*, Nature (London) **602**, 245 (2022).
[31] L. Yu, C. Wang, Y. Zhang, M. Sander, S. Ni, Z. Lu, S. Ma, Z. Wang, Z. Zhao, and H. Chen et al., *Evidence of a Hidden Flux Phase in The Topological Kagome Metal $CsV_3Sb_5$*. arXiv:2107.10714.
[32] W. Duan, Z. Nie, S. Luo, F. Yu, B. R. Ortiz, L. Yin, H. Su, F. Du, A. Wang, Y. Chen, X. Lu, J. Ying, S. D. Wilson, X. Chen, Y. Song, and H. Yuan, *Nodeless Superconductivity in the Kagome Metal $CsV_3Sb_5$*, Sci. China-Phys. Mech. Astron. **64**, 107462 (2021).
[33] W. A. Little and R. D. Parks, *Observation of Quantum Periodicity in the Transition Temperature of a Superconducting Cylinder*, Phys. Rev. Lett. **9**, 9 (1962).
[34] V. G. Kogan, J. R. Clem, and R. G. Mints, *Properties of mesoscopic superconducting thin-film rings: London approach*, Phys. Rev. B **69**, 064516 (2004).
[35] J. H. Han and A. Lee Patrick, *Understanding Resistance Oscillation in $CsV_3Sb_5$ Superconductor*, Phys. Rev. B **106**, 184515 (2022).
[36] V. B. Geshkenbein, A. I. Larkin, and A. Barone, *Vortices with Half Magnetic Flux Quanta in "Heavy-Fermion" Superconductors*, Phys. Rev. B **36**, 235 (1987).
[37] H.-Y. Kee, Y. B. Kim, and K. Maki, *Half-quantum Vortex and d-soliton in $Sr_2RuO_4$*, Phys. Rev. B **62**, R9275 (2000).
[38] M. Sigrist, T. M. Rice, and K. Ueda, *Low-field Magnetic Response of Complex Superconductors*, Phys. Rev. Lett. **63**, 1727 (1989).
[39] M. A. Rampp and J. Schmalian, *Integer and Fractionalized Vortex Lattices and Off-diagonal Long-range Order*, J. Phys. Commun. **6**, 055013 (2022).

[40] J. M. Kosterlitz and D. J. Thouless, *Ordering, metastability and phase transitions in two-dimensional systems*, J. Phys. C **6**, 1181 (1973).
[41] D. R. Nelson and B. I. Halperin, *Dislocation-mediated melting in two dimensions*, Phys. Rev. B **19**, 2457 (1979).
[42] A. P. Young, *Melting and the vector Coulomb gas in two dimensions*, Phys. Rev. B **19**, 1855 (1979).
[43] C. Mu, Q. Yin, Z. Tu, C. Gong, H. Lei, Z. Li, and J. Luo, *S-Wave Superconductivity in Kagome Metal $CsV_3Sb_5$ Revealed by $^{121/123}$Sb NQR and $^{51}$V NMR Measurements*, Chin. Phys. Lett. **38**, 077402 (2021).
[44] R. Gupta, D. Das, C. Mielke III, Z. Guguchia, T. Shiroka, C. Baines, M. Bartkowiak, H. Luetkens, R. Khasanov, Q. Yin, Z. Tu, G. Chunsheng, and H. Lei, *Microscopic evidence for anisotropic multigap superconductivity in the $CsV_3Sb_5$ kagome superconductor*, Npj Quantum Mater. **7**, 49 (2022).
[45] Z. Shan, P. K. Biswas, S. K. Ghosh, T. Tula, A. D. Hillier, D. Adroja, S. Cottrell, G.-H. Cao, Y. Liu, X. Xu, Y. Song, H. Yuan, and M. Smidma, *Muon Spin Relaxation Study of The Layered Kagome Superconductor $CsV_3Sb_5$*, Phys. Rev. Research **4**, 033145 (2022).
[46] F. Loder, A. P. Kampf, T. Kopp, J. Mannhart, C. W. Schneider, and Y. S. Barash, *Magnetic flux periodicity of h/e in superconducting loops*, Nat. Phys. **4**, 112 (2008).